\pdfoutput=1
\documentclass[11pt,twoside,a4paper,cmspaper,final,collab]{cms-tdr}
\def\svnVersion{0b224fe}\def\svnDate{2023/08/02}\def\cmsCernNoTag{CERN-EP-2022-168}\def\cmsCernDate{\today}\def\cmsMessage{Published in the Journal of High Energy Physics as \href{http://dx.doi.org/10.1007/JHEP07(2023)219}{\doi{10.1007/JHEP07(2023)219}.}}
\begin{document}\cmsNoteHeader{TOP-21-011}

\newlength\cmsTabSkip\setlength\cmsTabSkip{1ex}

\newcommand{\pp}{\ensuremath{\Pp\Pp}\xspace}
\newcommand{\ud}{\ensuremath{\PQu\PAQd}\xspace}
\newcommand{\du}{\ensuremath{\PQd\PAQu}\xspace}
\newcommand{\ttW}{\ensuremath{\ttbar\PW}\xspace}
\newcommand{\ttWp}{\ensuremath{\ttbar\PWp}\xspace}
\newcommand{\ttWm}{\ensuremath{\ttbar\PWm}\xspace}
\newcommand{\ttH}{\ensuremath{\ttbar\PH}\xspace}
\newcommand{\ttZ}{\ensuremath{\ttbar\PZ}\xspace}
\newcommand{\ttZG}{\ensuremath{\ttbar\PZ/\PGg^\ast}\xspace}
\newcommand{\tHq}{\ensuremath{\PQt\PH\PQq}\xspace}
\newcommand{\tZq}{\ensuremath{\PQt\PZ\PQq}\xspace}
\newcommand{\tHW}{\ensuremath{\PQt\PH\PW}\xspace}
\newcommand{\ttg}{\ensuremath{\ttbar\PGg}\xspace}
\newcommand{\ttVV}{\ensuremath{\ttbar\PV\PV}\xspace}
\newcommand{\tttt}{\ensuremath{\ttbar\ttbar}\xspace}
\newcommand{\ttV}{\ensuremath{\PQt\PAQt\PV}\xspace}
\newcommand{\Wjets}{\ensuremath{\PW\text{+jets}}\xspace}
\newcommand{\WZ}{\ensuremath{\PW\PZ}\xspace}
\newcommand{\ZZ}{\ensuremath{\PZ\PZ}\xspace}
\newcommand{\VVV}{\ensuremath{\PV\PV\PV}\xspace}

\newcommand{\TensorFlow}{\textsc{TensorFlow}\xspace}
\newcommand{\Keras}{\textsc{keras}\xspace}
\newcommand{\Adam}{\textsc{adam}\xspace}

\newcommand{\mll}{\ensuremath{m(\Pell\Pell)}\xspace}
\newcommand{\mlll}{\ensuremath{m(3\Pell)}\xspace}
\newcommand{\sigeta}{\ensuremath{\eta}\xspace}
\newcommand{\abseta}{\ensuremath{\abs{\eta}}\xspace}
\newcommand{\DPhi}{\ensuremath{\Delta\phi}\xspace}
\newcommand{\DEta}{\ensuremath{\Delta\eta}\xspace}
\newcommand{\dxy}{\ensuremath{d_{xy}}\xspace}
\newcommand{\dz}{\ensuremath{d_{z}}\xspace}
\newcommand{\mtfix}{\ensuremath{\mT^{\text{fixed}}}\xspace}
\newcommand{\ptfix}{\ensuremath{\pt^{\text{fixed}}}\xspace}
\newcommand{\DRmax}{\ensuremath{R_{\text{max}}}\xspace}
\newcommand{\DmllZ}{\ensuremath{\abs{\mll-\mZ}}\xspace}
\newcommand{\mZ}{\ensuremath{m_{\PZ}}\xspace}

\newcommand{\sigTTW}{\ensuremath{\sigma_{\ttW}}\xspace}
\newcommand{\sigTTWp}{\ensuremath{\sigma_{\ttWp}}\xspace}
\newcommand{\sigTTWm}{\ensuremath{\sigma_{\ttWm}}\xspace}
\newcommand{\RTTWpm}{\ensuremath{\sigma_{\ttWp}/\sigma_{\ttWm}}\xspace}

\newcommand{\fb}{\ensuremath{\unit{fb}}\xspace}

\newcommand{\ExpXsecTTW}{\ensuremath{868\pm40\stat\pm51\syst\fb}\xspace}
\newcommand{\ExpXsecTTWpNU}{\ensuremath{553\pm30\stat\pm30\syst}\xspace}
\newcommand{\ExpXsecTTWp}{\ensuremath{\ExpXsecTTWpNU\fb}\xspace}
\newcommand{\ExpXsecTTWm}{\ensuremath{343\pm26\stat\pm25\syst\fb}\xspace}
\newcommand{\ExpXsecTTWr}{\ensuremath{1.61\pm0.15\stat\,^{+0.07}_{-0.05}\syst}\xspace}
\newcommand{\PredXsecTTWnnlo}{\ensuremath{592\,^{+155}_{-97}\thy\fb}\xspace}

\newcommand{\sqrts}[1][13]{\ensuremath{\sqrt{s}=#1\TeV}\xspace}
\newcommand{\il}{\ensuremath{138\fbinv}\xspace}
\newcommand{\alpEW}{\ensuremath{\alpha}\xspace}

\newcommand{\PQj}{\HepParticle{j}{}{}\xspace}
\newcommand{\Pellp}{\HepParticle{\Pell}{}{+}\xspace}
\newcommand{\Pellm}{\HepParticle{\Pell}{}{-}\xspace}
\newcommand{\Pellpm}{\HepParticle{\Pell}{}{\pm}\xspace}
\newcommand{\Pellmp}{\HepParticle{\Pell}{}{\mp}\xspace}

\newcommand{\ttl}{\ensuremath{f}\xspace}
\newcommand{\evw}{\ensuremath{w}\xspace}
\newcommand{\num}{\ensuremath{n}\xspace}
\newcommand{\Npass}{\ensuremath{N_{\text{pass}}}\xspace}
\newcommand{\Nfail}{\ensuremath{N_{\text{fail}}}\xspace}

\newcommand{\NLONNLL}{\ensuremath{\text{NLO\,+\,NNLL}}\xspace}
\newcommand{\NLOFXFX}{\ensuremath{\text{NLO\,+\,FxFx}}\xspace}

\newcommand{\muR}{\ensuremath{\mu_{\mathrm{R}}}\xspace}
\newcommand{\muF}{\ensuremath{\mu_{\mathrm{F}}}\xspace}

\cmsNoteHeader{TOP-21-011}
\title{Measurement of the cross section of top quark-antiquark pair production in association with a \PW boson in proton-proton collisions at \texorpdfstring{\sqrts}{sqrt(s)=13 TeV}}

\date{\today}

\abstract{The production of a top quark-antiquark pair in association with a
\PW boson (\ttW) is measured in proton-proton collisions at a center-of-mass
energy of 13\TeV.
The analyzed data was recorded by the CMS experiment at the CERN LHC and
corresponds to an integrated luminosity of \il.
Events with two or three leptons (electrons and muons) and additional jets are
selected.
In events with two leptons, a multiclass neural network is used to distinguish
between the signal and background processes.
Events with three leptons are categorized based on the number of jets and of
jets originating from \PQb quark hadronization, and the lepton charges.
The inclusive \ttW production cross section in the full phase space is measured
to be \ExpXsecTTW.
The \ttWp and \ttWm cross sections are also measured as \ExpXsecTTWpNU
and \ExpXsecTTWm, respectively, and the corresponding ratio of the two cross
sections is found to be \ExpXsecTTWr.
The measured cross sections are larger than but consistent with the standard
model predictions within two standard deviations, and represent the most precise
measurement of these cross sections to date.}

\hypersetup{
pdfauthor={CMS Collaboration},
pdftitle={Measurement of the cross section of top quark-antiquark pair production in association with a W boson in proton-proton collisions at sqrt(s)=13 TeV},
pdfsubject={CMS},
pdfkeywords={CMS, top quark, electroweak production}}

\maketitle

\section{Introduction}

\begin{figure}[h!]
\centering
\includegraphics[width=0.4\textwidth]{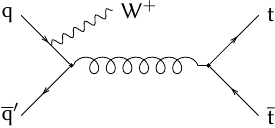}%
\hspace{0.1\textwidth}%
\includegraphics[width=0.4\textwidth]{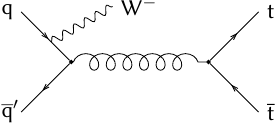} \\[0.04\textwidth]
\raisebox{0.034\textwidth}{\includegraphics[width=0.4\textwidth]{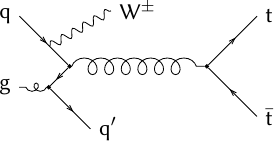}}%
\hspace{0.1\textwidth}%
\makebox[0.4\textwidth]{\includegraphics[width=0.325\textwidth]{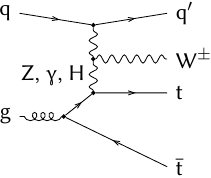}}%
\caption{%
    Representative Feynman diagrams of \ttW production at LO (upper row) and NLO
    (lower row).
}
\label{fig:feynman}
\end{figure}

The production of a top quark-antiquark pair (\ttbar) in association with a
vector boson (\PW and \PZ bosons and photons, referred to collectively as \PV)
in proton-proton (\pp) collisions provides an experimental probe that is
sensitive to the electroweak (EW) couplings of the top quark.
These \ttV processes can receive sizable contributions from phenomena beyond the
standard model (SM) of particle physics~\cite{BessidskaiaBylund:2016jvp,
Dror:2015nkp, Buckley:2015lku}.
The production process of \ttbar in association with a \PW boson (\ttW) is
different from the \ttV processes with neutral bosons (\ttZ and \ttg) since a
\PW boson cannot be radiated from the final-state top quarks.
The production proceeds at leading order (LO) in quantum chromodynamics (QCD)
and EW couplings from quark-antiquark initial states.
At next-to-LO (NLO), additional contributions arise from (anti)quark-gluon
initial states.
These dominant production modes are illustrated with representative Feynman
diagrams in Fig.~\ref{fig:feynman}.

The gluon-gluon initial states are suppressed in \ttW production, leading to a
sizable difference between production rates for the \ttWp and \ttWm processes
because of differences in the parton distribution functions (PDFs) of the
valence and sea quarks in the proton.
Specifically, for the main LO production channels $\ud\to\ttWp$ and
$\du\to\ttWm$, the \ud initial state has about twice the abundance
of the \du initial state and typically a larger longitudinal quark momentum,
resulting in a noticeable charge asymmetry~\cite{Maltoni:2014zpa}.
Compared to the other \ttV processes, the \ttW process receives large
corrections from the EW production diagrams at NLO, representing an important
challenge in the calculation of the uncertainties for this
process~\cite{Frederix:2017wme}.
Therefore, precise cross section measurements are crucial inputs to further
improve the calculations.

At the CERN LHC, \ttW production has been measured at
$\sqrt{s}=7$~\cite{CMS:TOP-12-014}, 8~\cite{ATLAS:2015qtq, CMS:TOP-14-021}, and
13\TeV~\cite{ATLAS:2016wgc, CMS:TOP-17-005, ATLAS:2019fwo}.
In recent measurements of \ttbar production in association with a Higgs boson
(\ttH)~\cite{CMS:HIG-19-008} and the production of four top quarks
(\tttt)~\cite{CMS:TOP-18-003, ATLAS:2020hpj}, a tension was observed between the
observed and predicted yields for the \ttW process, which represents one of the
dominant backgrounds in these results.
It has been shown that the observed difference can largely be attributed to
missing NLO corrections in theoretical calculations for this
process~\cite{Broggio:2019ewu, Kulesza:2020nfh, Bevilacqua:2020srb,
vonBuddenbrock:2020ter, Frederix:2020jzp, FebresCordero:2021kcc,
Frederix:2021agh, Bevilacqua:2021tzp}.

This paper presents a measurement of the inclusive \ttW production cross section
in the full phase space, based on an integrated luminosity of \il of \pp
collision data collected with the CMS detector at \sqrts between 2016 and 2018.
Two decay channels of the \ttW system are analyzed, with either two same-sign or
three charged leptons (any combination of electrons and muons, denoted as
\Pell).
The dominant background contributions arise from the presence of ``nonprompt''
leptons, which are leptons originating from hadron decays, as well as jets or
hadrons that are misidentified as leptons.
Suppression of these background events relies on multivariate analysis (MVA)
techniques, which are optimized to distinguish between prompt leptons
from the decay of \PW bosons and nonprompt leptons.
Additional background contributions in final states with electrons arise from
the misreconstruction of the electron charge and from the production of
electrons in photon conversions.
The dominant background contributions with prompt leptons are \ttZ and \ttH
production, and the associated production of a \PW and \PZ boson (\WZ).
The \ttW cross section is extracted from a profile likelihood fit to the data.

{\tolerance=800
This analysis supersedes the CMS measurement presented in
Ref.~\cite{CMS:TOP-17-005}, which used data recorded in 2016, corresponding to
an integrated luminosity of 35.9\fbinv.
In addition to the larger data set, the sensitivity is improved through the use
of new machine-learning-based lepton and \PQb hadron identification techniques.
A more rigorous event categorization and improved background prediction
techniques are also employed.
\par}

The paper is organized as follows:
Section~\ref{sec:detector} gives a description of the CMS detector.
In Section~\ref{sec:datasets}, the data set and simulated samples are described.
In Section~\ref{sec:reconstruction}, the event reconstruction is defined.
The event selection and analysis strategy are discussed in
Section~\ref{sec:eventselection}, followed by a description of the estimation of
background processes in Section~\ref{sec:backgrounds}.
The main sources of experimental and theoretical systematic uncertainties are
presented in Section~\ref{sec:systematics}.
Finally, the results and a summary are given in Sections~\ref{sec:results}
and~\ref{sec:summary}, respectively.
Tabulated results are provided in the HEPData record for this
analysis~\cite{hepdata}.

\section{The CMS detector}
\label{sec:detector}

The central feature of the CMS apparatus is a superconducting solenoid of
6\unit{m} internal diameter, providing a magnetic field of 3.8\unit{T}.
Within the solenoid volume are a silicon pixel and strip tracker, a lead
tungstate crystal electromagnetic calorimeter (ECAL), and a brass and
scintillator hadron calorimeter (HCAL), each composed of a barrel and two endcap
sections.
Forward calorimeters extend the pseudorapidity (\sigeta) coverage provided by the
barrel and endcap detectors.
Muons are detected in gas-ionization chambers embedded in the steel flux-return
yoke outside the solenoid.
A more detailed description of the CMS detector, together with a definition of
the coordinate system used and the relevant kinematic variables, can be found in
Ref.~\cite{CMS:Detector-2008}.

Events of interest are selected using a two-tiered trigger system.
The first level (L1), composed of custom hardware processors, uses information
from the calorimeters and muon detectors to select events at a rate of around
100\unit{kHz} within a fixed latency of about 4\mus~\cite{CMS:TRG-17-001}.
The second level, known as the high-level trigger, consists of a farm of
processors running a version of the full event reconstruction software optimized
for fast processing, and reduces the event rate to around 1\unit{kHz} before
data storage~\cite{CMS:TRG-12-001}.

\section{Data and simulated samples}
\label{sec:datasets}

The \pp collision data used for this measurement have been recorded with a set
of triggers that require the presence of one, two, or three reconstructed
electrons or muons.
The trigger thresholds depend on the data-taking year and on the number and
type of leptons required by the trigger.
The overall trigger efficiency after the analysis selection is above 80\%.
Simulated event samples are used to evaluate the signal selection efficiency,
validate the background estimates with nonprompt leptons, predict the background
contributions with prompt leptons, and train MVA classifiers.
Three separate sets of simulated event samples are used, corresponding to the
conditions of the three data-taking years.

For the simulation of the \ttW signal process at NLO accuracy, terms of the
order $\alpS\alpEW^3$ and $\alpS^3\alpEW$ are relevant~\cite{Dror:2015nkp,
Frederix:2017wme, Bevilacqua:2020pzy, Frederix:2020jzp}, where \alpS and \alpEW
are the strong coupling and fine-structure constants, respectively.
Events are simulated with the \MGvATNLO v2.6.0~\cite{EXT:MG5aMCatNLO-2014}
program, taking LO and NLO terms of order $\alpEW^3$ and $\alpS\alpEW^3$ (``EWK
terms''), as well as $\alpS^2\alpEW$ and $\alpS^3\alpEW$ (``QCD terms'') into
account.
The theoretical \ttW cross section is \PredXsecTTWnnlo, which is obtained at NLO
in QCD and EW effects, and includes QCD corrections at
next-to-next-to-leading-logarithmic (NNLL) accuracy~\cite{Kulesza:2020nfh}.
The uncertainty in the prediction accounts for matrix-element (ME) scale
variations and the PDF choice.
We use a cross section value of 49.1\fb, as evaluated with the generator, to
normalize the event sample simulated for the EWK terms in the ME calculation,
and the remainder of the full NLO+NNLL cross section prediction for the event
sample simulated for the QCD terms.

The \MGvATNLO program is also used to simulate background events at NLO accuracy
for the production of \ttbar or a single top quark in association with a vector
boson ($\ttbar\Pellp\Pellm$ with $\mll>1\GeV$, referred to as ``\ttZG'', \ttg,
\tZq, and $\PQt\PGg$), as well as of \tttt, Drell--Yan (DY), \VVV, \WZ, and
$\PV\PH$ production.
Background processes for single top quark production in association with a Higgs
boson (\tHq and \tHW) and for \ttVV production are simulated with \MGvATNLO at
LO accuracy.
The \POWHEG v2~\cite{EXT:Powheg-2004, EXT:Powheg-2007, EXT:Powheg-2010} program
is used to simulate events for the production of two \PW ($\PW\PW$) or two \PZ
bosons (\ZZ), \ttbar, and \ttH at NLO precision.
The background from SM events composed uniquely of jets produced through the
strong interaction, referred to as QCD multijet events, is simulated at LO with
\PYTHIA v8.226 (v8.230)~\cite{EXT:Pythia-2015} in 2016 (2017--2018), and is used
for the validation of background predictions associated with processes involving
nonprompt leptons.

All samples are processed with the \PYTHIA parton shower (PS) simulation.
In the event simulation using \MGvATNLO, the FxFx~\cite{EXT:FxFx-2012}
(MLM~\cite{EXT:Merging-2008}) merging scheme is used for NLO (LO) samples, in
order to avoid possible double counting between the ME and PS calculations.
For the \ttW signal sample, a merging scale of 42\GeV is used to define the
limit at which the PS effects are matched to the ME calculation, and the values
of the renormalization and factorization scales, \muR and \muF, are set
dynamically to correspond to the characteristic scale of the hard process,
depending on the kinematic properties of the event after the FxFx merging
prescription as described in Ref.~\cite{EXT:FxFx-2012}.
The propagation of simulated particles through the CMS detector and modeling of
the detector response is performed using the \GEANTfour
toolkit~\cite{EXT:GEANT4-2002}.
The simulated samples include inelastic \pp interactions in the same or nearby
bunch crossing (pileup), assuming a value of 69.2\unit{mb} for the total
inelastic \pp cross section~\cite{CMS:FWD-11-002, CMS:FWD-11-001}.

In the simulation used for the 2017--2018 data, the
NNPDF3.1~\cite{EXT:NNPDF-2017} sets are used for the PDFs, and the underlying
event is modeled with the CP5 tuning parameters~\cite{Skands:2014pea,
CMS:GEN-17-001}.
The 2016 samples for most processes are simulated using the same PDF sets and PS
tune.
For some background processes, however, the 2016 samples are modeled using the
NNPDF3.0~\cite{EXT:NNPDF-2015} PDF sets, and with the
CUETP8M1~\cite{CMS:GEN-14-001} or CUETP8M2T4~\cite{CMS:TOP-16-021} tunes.

\section{Event reconstruction}
\label{sec:reconstruction}

The particle-flow (PF) algorithm~\cite{CMS:PRF-14-001} aims to reconstruct and
identify each individual particle in an event, with an optimized combination of
information from the various elements of the CMS detector.
The energy of photons is obtained from the ECAL measurement.
The energy of electrons is determined from a combination of the track momentum
at the main interaction vertex, the corresponding ECAL cluster energy, and the
energy sum of all bremsstrahlung photons attached to the track.
The primary interaction vertex is taken to be the vertex corresponding to the
hardest scattering in the event, evaluated using tracking information alone, as
described in Section 9.4.1 of Ref.~\cite{CMS:TDR-15-02}.
The energy of muons is obtained from the curvature of the corresponding track.
The energy of charged hadrons is determined from a combination of their momentum
measured in the tracker and the matching ECAL and HCAL energy deposits,
corrected for the response function of the calorimeters to hadronic showers.
Finally, the energy of neutral hadrons is obtained from the corresponding
corrected ECAL and HCAL energies.

Electrons are identified with an MVA discriminant~\cite{CMS:EGM-17-001} and are
required to have transverse momentum $\pt>10\GeV$ and $\abseta<2.5$.
Electrons that are reconstructed within the transition region between the barrel
and the endcap sections of the detector within $1.479<\abseta<1.566$, as well as
those associated with tracks with more than one missing hit in the tracking
system, are excluded from the analysis.

Muons are identified as tracks in the central tracker consistent with either a
track or several hits in the muon system, and associated with calorimeter
deposits compatible with the muon hypothesis.
Reconstructed muons must have $\pt>10\GeV$ and $\abseta<2.4$, and fulfill
selection criteria on the geometrical matching of the tracks reconstructed in
the silicon tracker and the muon system, and on the quality of the global
fit~\cite{CMS:MUO-16-001}.

The ``isolation'' of a lepton is defined as the sum of the \pt of all
reconstructed PF particles inside a cone of
$\DR=\sqrt{\smash[b]{(\DEta)^2+(\DPhi)^2}}<\DRmax$ (where \DEta and \DPhi are
the \sigeta and the azimuthal angle differences, respectively) around the
direction of the lepton, excluding the lepton itself.
We use a variable cone size that depends on the lepton \pt, defined as
$\DRmax=10/\!\min[\max(\pt,\,50),\,200]$ (where \pt is given in \GeVns).
The relative isolation, \ie, the isolation divided by the lepton \pt, is
required to be less than 0.4.
Any electron is furthermore required to be separated from any muon by
$\DR>0.05$.

The impact parameter of a lepton track with respect to the primary interaction
vertex, with components in the transverse (\dxy) and longitudinal (\dz) planes
relative to the direction of colliding proton beams, must satisfy
$\abs{\dxy}<0.05\cm$ and $\abs{\dz}<0.1\cm$, respectively.
The impact parameter significance of the lepton track, defined as the absolute
value of the ratio between the impact parameter value and its uncertainty, must
be less than 8.

Two additional sets of lepton selection criteria, referred to as ``loose'' and
``tight'' in the following, are used to suppress background events associated
with the presence of nonprompt leptons.
The tight definition includes a requirement on a dedicated lepton MVA
discriminant optimized for the selection of leptons in top quark events, which
has been used in Refs.~\cite{CMS:TOP-20-010, CMS:TOP-21-004} and is based on the
earlier developments in Refs.~\cite{CMS:SUS-15-008, CMS:TOP-18-008}.
The prompt-lepton selection efficiency of this discriminant is evaluated in
simulated event samples to be about 95\% for leptons with $\pt>25\GeV$, and
50--70\% for leptons with $10<\pt<25\GeV$, depending on \abseta.
The loose definition includes all tight leptons and additionally leptons that
fail the lepton MVA requirement but instead pass loose requirements on two
properties of the reconstructed jet closest to  the lepton, which are also input
variables to the lepton MVA.
Additionally, loose and tight electrons must not be associated with a secondary
vertex that is consistent with a photon conversion~\cite{CMS:EGM-17-001}, must
have a charge measurement consistent among the three independent measurements
described in Ref.~\cite{CMS:EGM-13-001}, and fulfill additional criteria based
on the shape of the electromagnetic shower and the consistency between the track
and the shower~\cite{CMS:EGM-17-001}.

In each event, jets are clustered from the reconstructed PF particles with the
anti-\kt algorithm~\cite{EXT:AK-2008, EXT:FastJet-2012} using a distance
parameter of 0.4.
The jet momentum is determined as the vector sum of the momenta of all particles
in the jet, and is found from simulation to be, on average, within 5--10\% of
the true momentum over the entire \pt spectrum and detector acceptance.
Additional tracks and calorimetric energy depositions can come from pileup,
increasing the apparent jet momentum.
To mitigate this effect, tracks identified as originating from pileup vertices
are discarded and an offset correction is applied to correct for remaining
contributions~\cite{CMS:JME-18-001}.
Jet energy corrections are derived from simulation studies so that the average
measured energy of jets becomes identical to that of particle-level jets.
In situ measurements of the momentum balance in dijet, photon+jet, \PZ{}+jet,
and QCD multijet events are used to determine any residual differences between
the jet energy scale in data and simulation, and appropriate corrections are
applied~\cite{CMS:JME-13-004}.
Additional selection criteria are applied to each jet to remove jets potentially
dominated by instrumental effects or reconstruction
failures~\cite{CMS:JME-18-001}.
Jets with $\pt>25\GeV$ and $\abseta<2.4$ are considered in the analysis, and
those including a loose lepton among their PF constituents inside a cone of
$\DR<0.4$ are removed.

The DeepJet algorithm~\cite{CMS:BTV-16-002, EXT:DeepJet-2020, CMS:DP-2018-058}
is used to identify jets arising from the hadronization of \PQb hadrons
(\PQb-tagged jets).
The ``loose'' and ``medium'' \PQb tagging selection requirements in this
analysis correspond to selection efficiencies of about 90 and 85\%,
respectively, for \PQb quark jets with $\pt>30\GeV$ as estimated in simulated
\ttbar events.
For \PQc quark jets (light quark and gluon jets), misidentification rates of
50 and 15\% (10 and 1\%) are obtained.

The missing transverse momentum vector \ptvecmiss is computed as the negative
vector sum of the \ptvec of all reconstructed PF particles in an event, and its
magnitude is denoted as \ptmiss~\cite{CMS:JME-17-001}.
The value of \ptvecmiss is modified to account for corrections to the energy
scale of the reconstructed jets in the event.

\section{Event selection and analysis strategy}
\label{sec:eventselection}

The measurement targets the \ttW signal process in which the \PW boson decays to
a lepton (electron or muon) and a neutrino, and the final state of the \ttbar
system includes either lepton+jets or dileptons.
We select events containing two or three leptons.
The dilepton and trilepton channels are analyzed separately, and combined for
the final cross section measurement.
To ensure that orthogonal event samples are used for the analyses of the two
channels, only events with exactly two (three) loose leptons are considered in
the dilepton (trilepton) channel.
The signal regions of each channel are then defined by requiring that all of the
loose leptons also pass the tight selection, which have further requirements
discussed in the remainder of this section.
Events with at least one loose-not-tight lepton are used in each channel for the
estimation of the nonprompt-lepton backgrounds, as discussed in
Section~\ref{sec:backgrounds}.

\textbf{Same-sign dilepton channel:}\quad
Events in the dilepton channel are required to have exactly two tight leptons
and no additional loose leptons.
We require that the two leptons have the same sign electric charge, which
significantly reduces background contributions from SM processes with two
opposite-sign prompt leptons (\eg, \ttbar and DY production).
The highest \pt (leading) lepton is required to have $\pt>30\GeV$ if it is an
electron, or $\pt>25\GeV$ if it is a muon.
The second-highest \pt (subleading) lepton is required to have $\pt>20\GeV$,
independent of flavor.
To reduce background contributions from low-mass resonances and photon
conversions, the dilepton invariant mass \mll is required to be larger than
30\GeV, and the separation between the leptons must be $\DR>0.4$.
Events with two electrons and $\abs{m(\Pe\Pe)-\mZ}<15\GeV$ are removed, where
\mZ is the world-average \PZ boson mass~\cite{EXT:PDG-2020}.
The events are also required to have $\ptmiss>30\GeV$ and at least two jets, out
of which at least two pass the loose \PQb tagging selection or at least one
passes the medium \PQb tagging selection, which reduces background contributions
from DY and \ZZ production.

A multiclass neural network (NN) classifier is used to distinguish between
events from the signal \ttW process and those from three classes of background
processes: \ttZG and \ttH as a combined category, \ttg, and backgrounds with
nonprompt leptons.
The classifier is trained with the simulated \ttW, \ttZG, \ttH, and \ttg samples,
and with simulated \ttbar events as a source of nonprompt-lepton background
events.

The NN has two fully connected hidden layers with 128 and 64 rectified linear
units~\cite{Relu:2011}, respectively, and an output layer with four nodes
corresponding to the four categories defined above.
The training is performed using the \TensorFlow package~\cite{Tensorflow:2016}
with the \Keras interface~\cite{Keras:2015} and the \Adam
optimizer~\cite{Adam:2014}.
The training phase aims at reducing the cross-entropy loss
function~\cite{Crossentropy:2016}, and potential overtraining is mitigated using
dropout~\cite{Dropout:2014}.

The value of each output node can be interpreted as the probability for a given
event to belong to that corresponding category.
While we only use the distribution of the \ttW output node in the fit to extract
the \ttW cross section, the inclusion of the three separate background classes
with significantly different kinematic properties in a multiclass NN leads to a
better discriminating power between signal and background events compared to a
binary classification approach, even though only a single output node is used in
the fit.
The final NN discriminant that is used in the fit is obtained from a monotonic
transformation of the \ttW output node value, which is designed to give an
approximately linearly decreasing distribution of the total expected background.

\begin{figure}[p!]
\centering
\includegraphics[width=0.42\textwidth]{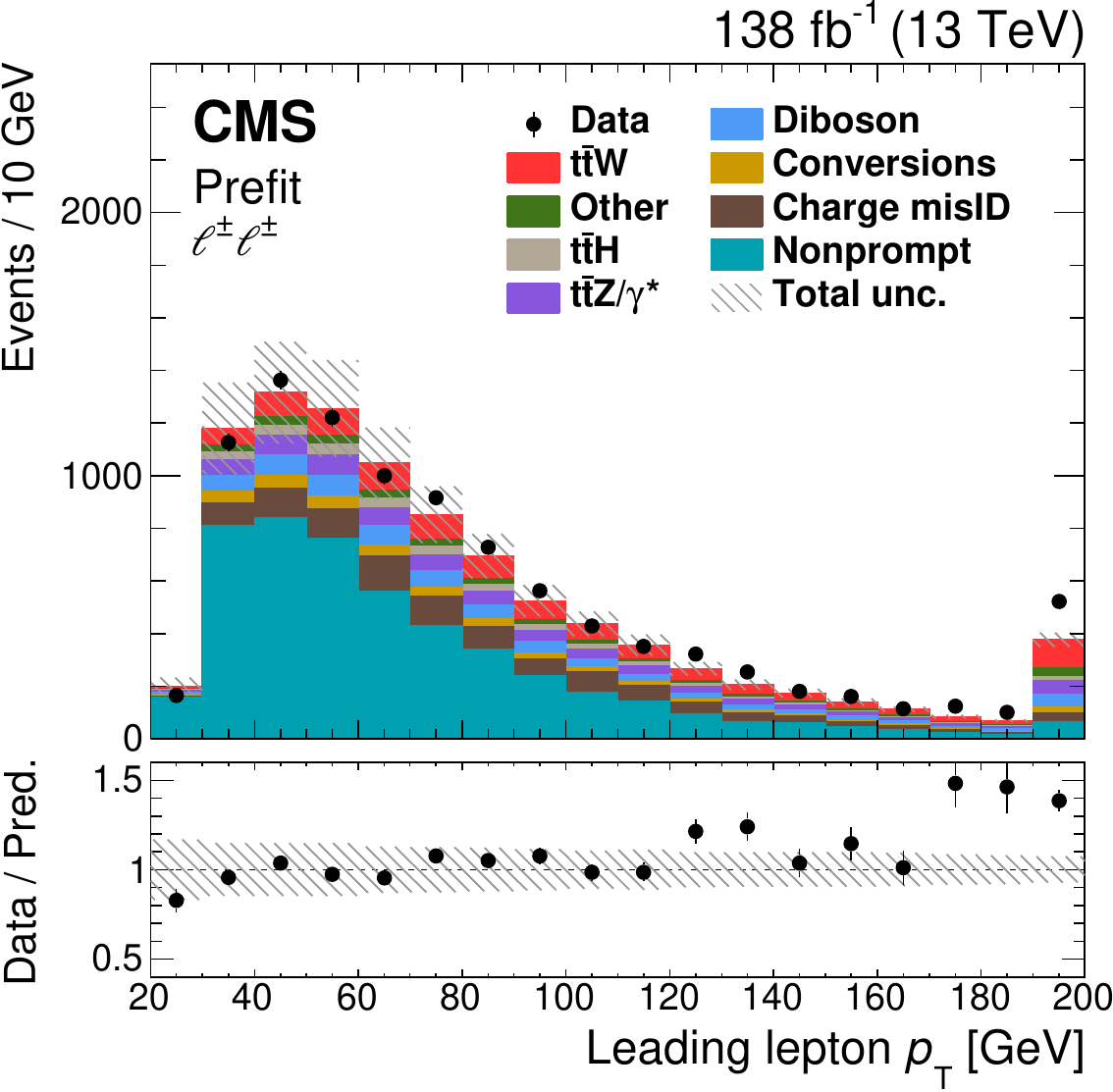}%
\hspace*{0.05\textwidth}%
\includegraphics[width=0.42\textwidth]{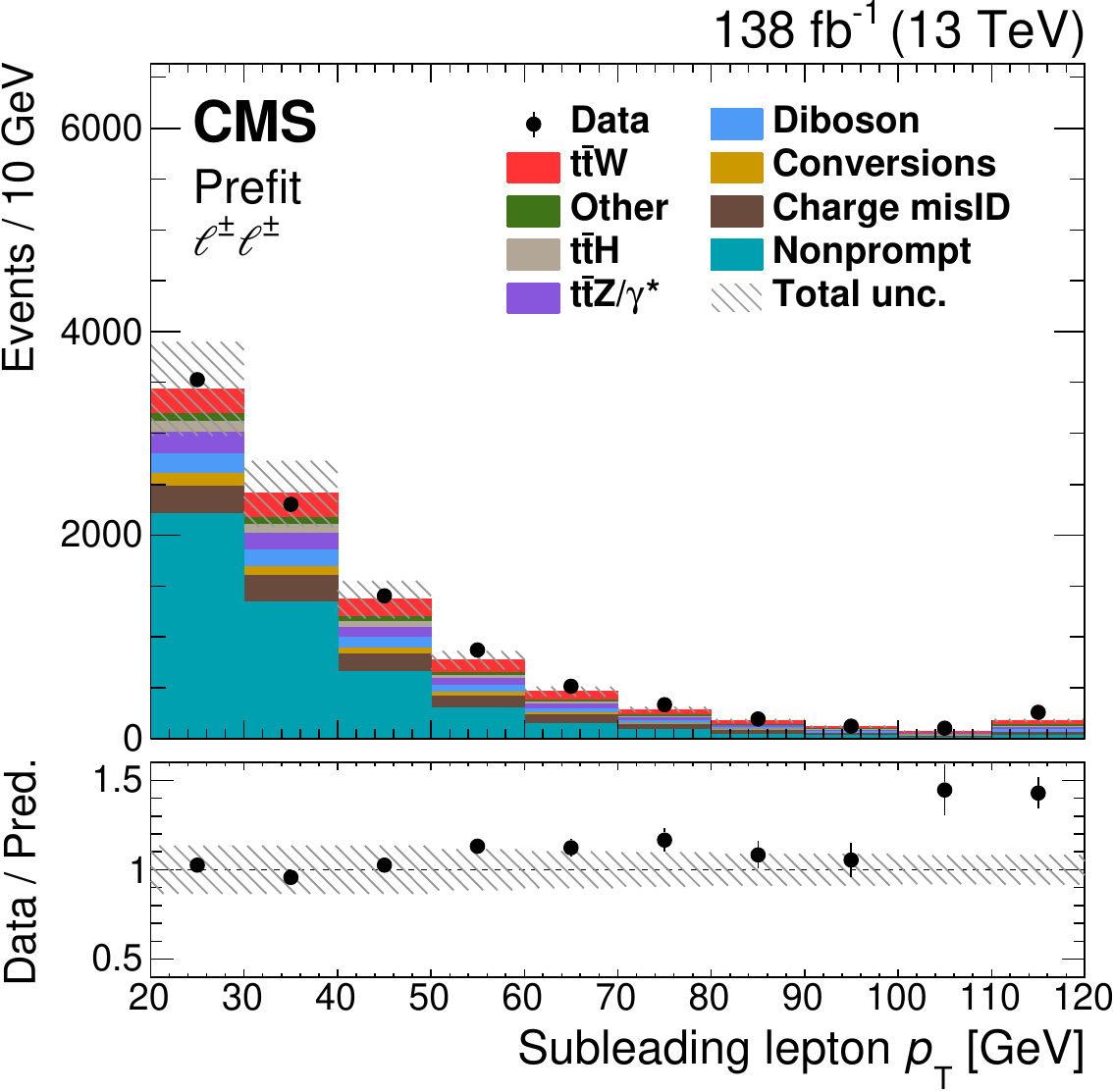} \\[1ex]
\includegraphics[width=0.42\textwidth]{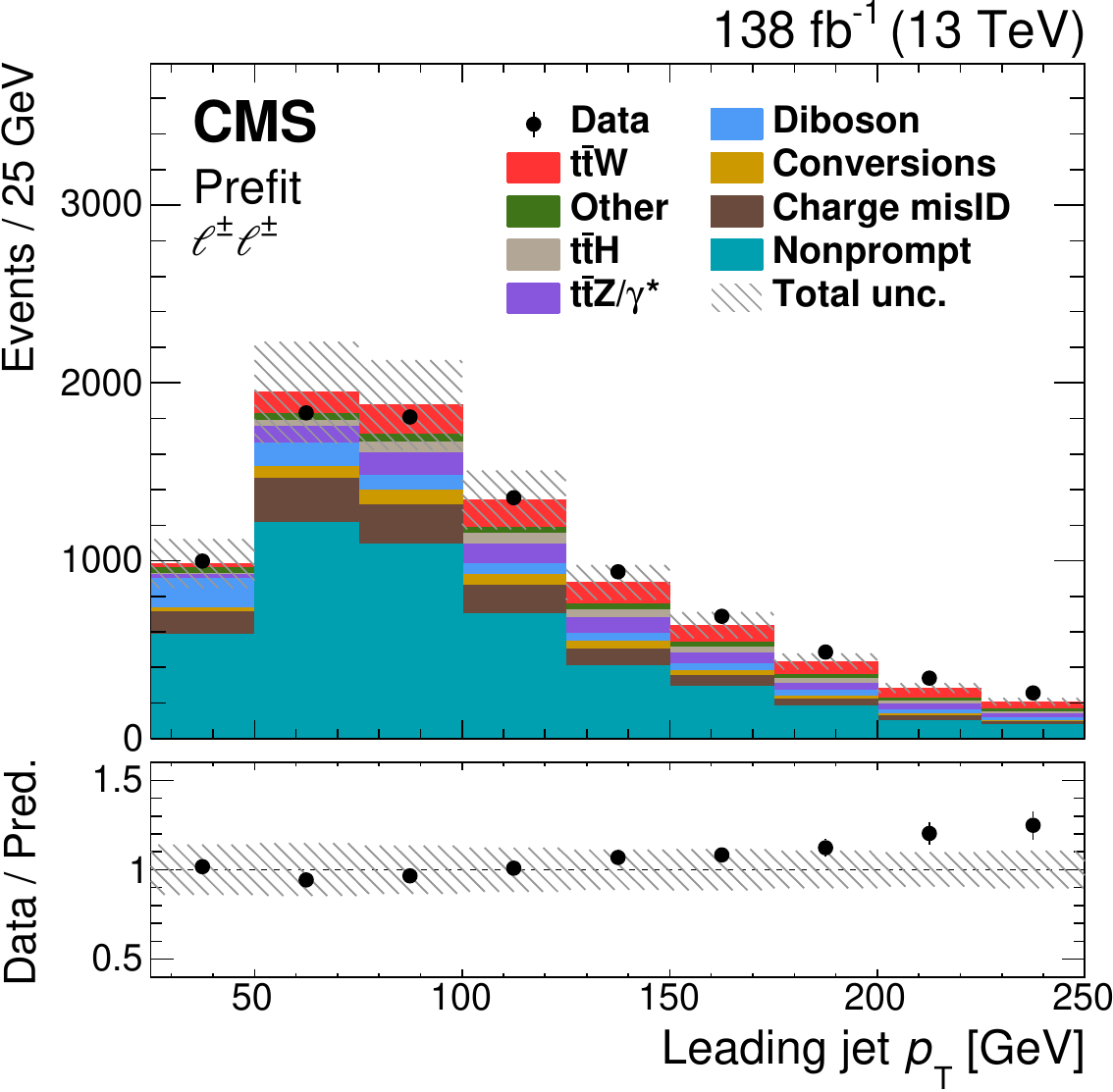}%
\hspace*{0.05\textwidth}%
\includegraphics[width=0.42\textwidth]{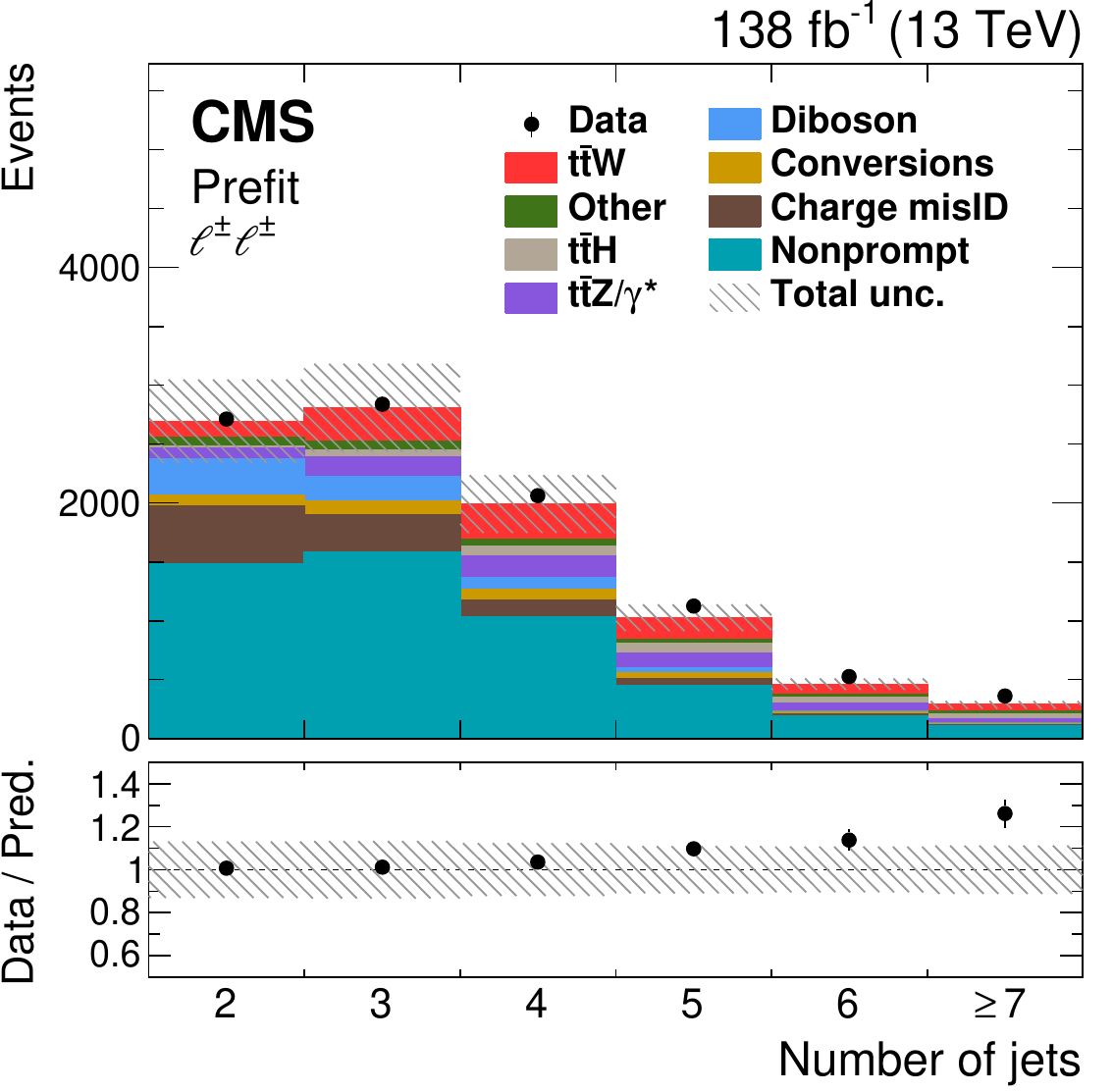} \\[1ex]
\includegraphics[width=0.42\textwidth]{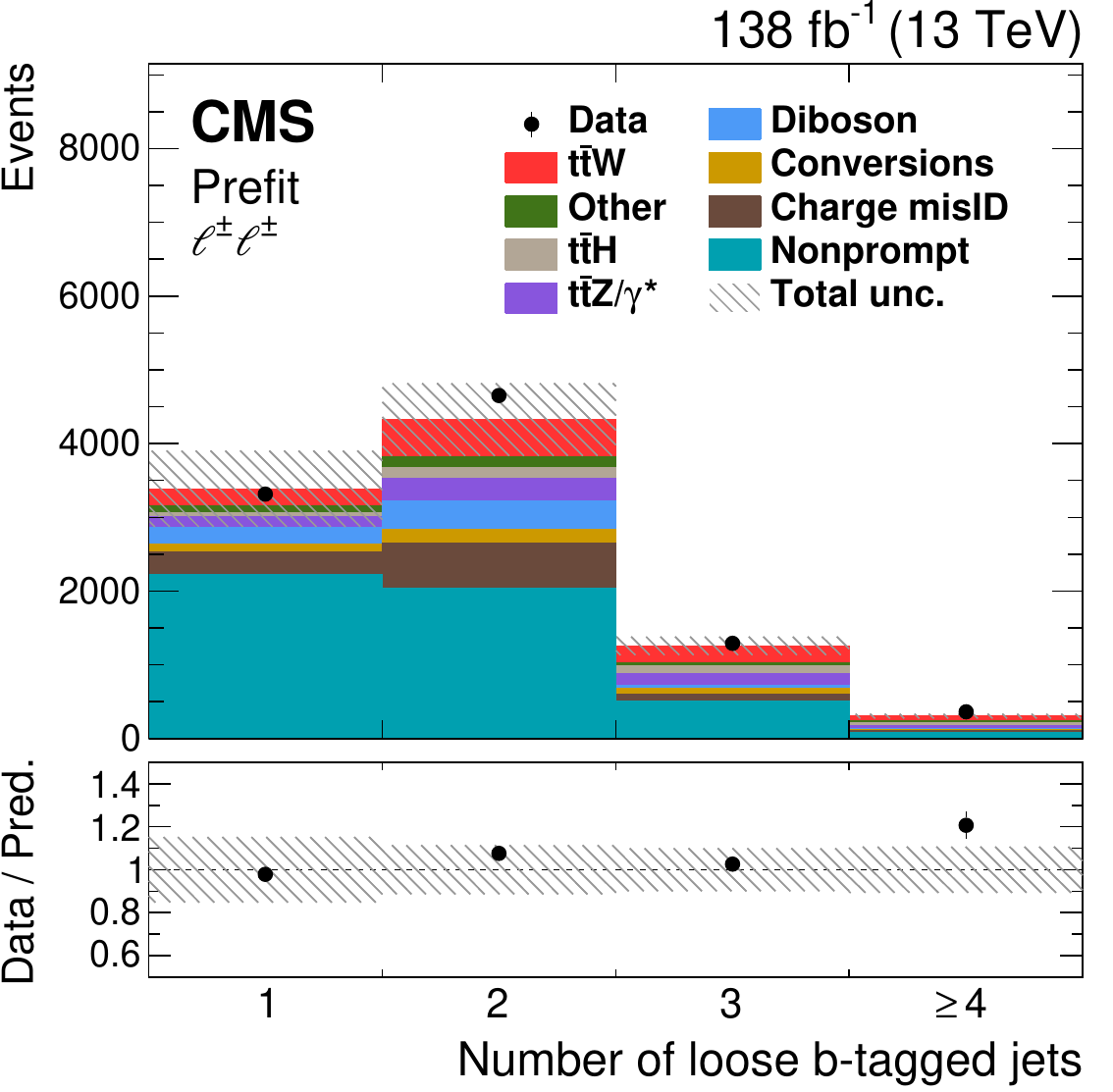}%
\hspace*{0.05\textwidth}%
\includegraphics[width=0.42\textwidth]{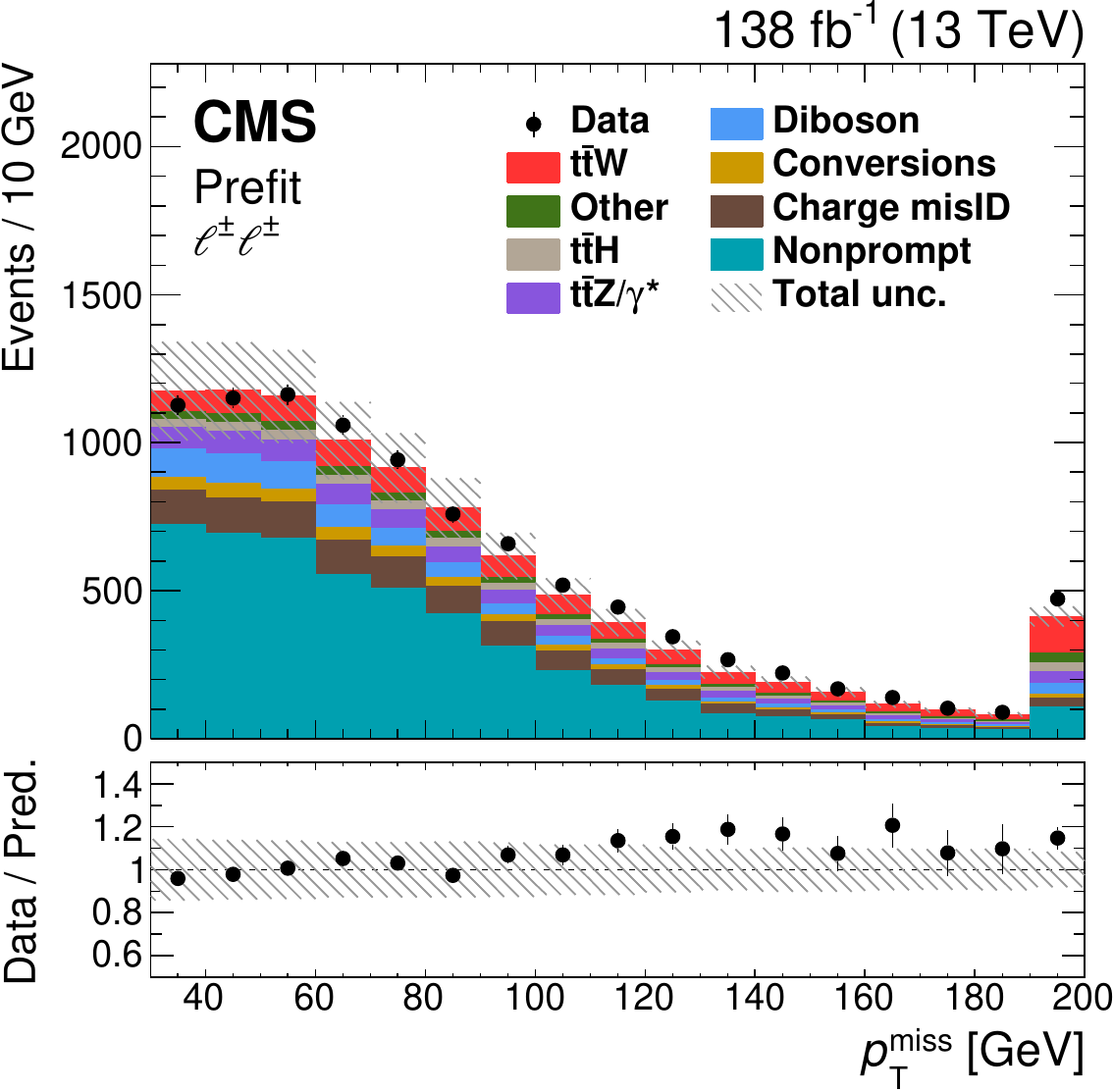}%
\caption{%
    Comparison of the number of observed (points) and predicted (colored
    histograms) events in the same-sign dilepton signal region.
    The distributions of the leading (upper left) and subleading (upper right)
    lepton \pt, the leading jet \pt (middle left), the number of jets (middle
    right), the number of loose \PQb-tagged jets (lower left), and \ptmiss
    (lower right) are displayed.
    The predictions are shown before the fit to data (``Prefit'').
    The vertical bars on the points represent the statistical uncertainties in
    the data, and the hatched bands the systematic uncertainty in the
    predictions.
    The last bins include the overflow contributions.
    In the lower panels, the ratio of the event yields in data to the overall
    sum of the predictions is presented.
}
\label{fig:sel2l}
\end{figure}

The input variables include several kinematic observables associated with the
reconstructed leptons and jets, such as \pt, $\phi$, \DR, and invariant masses
calculated for several combinations of leptons and jets.
Further input variables are the jet and \PQb-tagged jet multiplicities, the \PQb
tagging discriminant values of the selected jets, information about the flavor
and charge of the leptons, and \ptmiss.
In Fig.~\ref{fig:sel2l}, six of the input variables that provide the best
discrimination between the four classes of events are shown.
Overall, a reasonably good agreement between data and background predictions is
observed, with some disagreement in the tails of the distributions.
These discrepancies are reduced once the results from the fit are incorporated,
leading to an improved agreement between data and prediction.
We have validated that the agreement between data and simulation for all the
input variables is reasonable.
Additional validations have been performed in the control regions defined in
Section~\ref{sec:backgrounds}.

\textbf{Trilepton channel:}\quad
Events in the trilepton channel are required to have exactly three tight
leptons, with $\pt>25\GeV$ for the leading lepton and $\pt>15\GeV$ for the other
two leptons.
It is required that $\mll>12\GeV$ for each pair of leptons to remove background
contributions from low-mass resonances, and $\DmllZ>10\GeV$ for each pair of
leptons of same flavor and opposite sign to remove DY contributions.
The sum of charges of the three leptons is required to be $\pm1$, and events
must contain at least two jets, out of which at least one passes the medium \PQb
tagging selection.

\begin{figure}[ht!]
\centering
\includegraphics[width=0.6\textwidth]{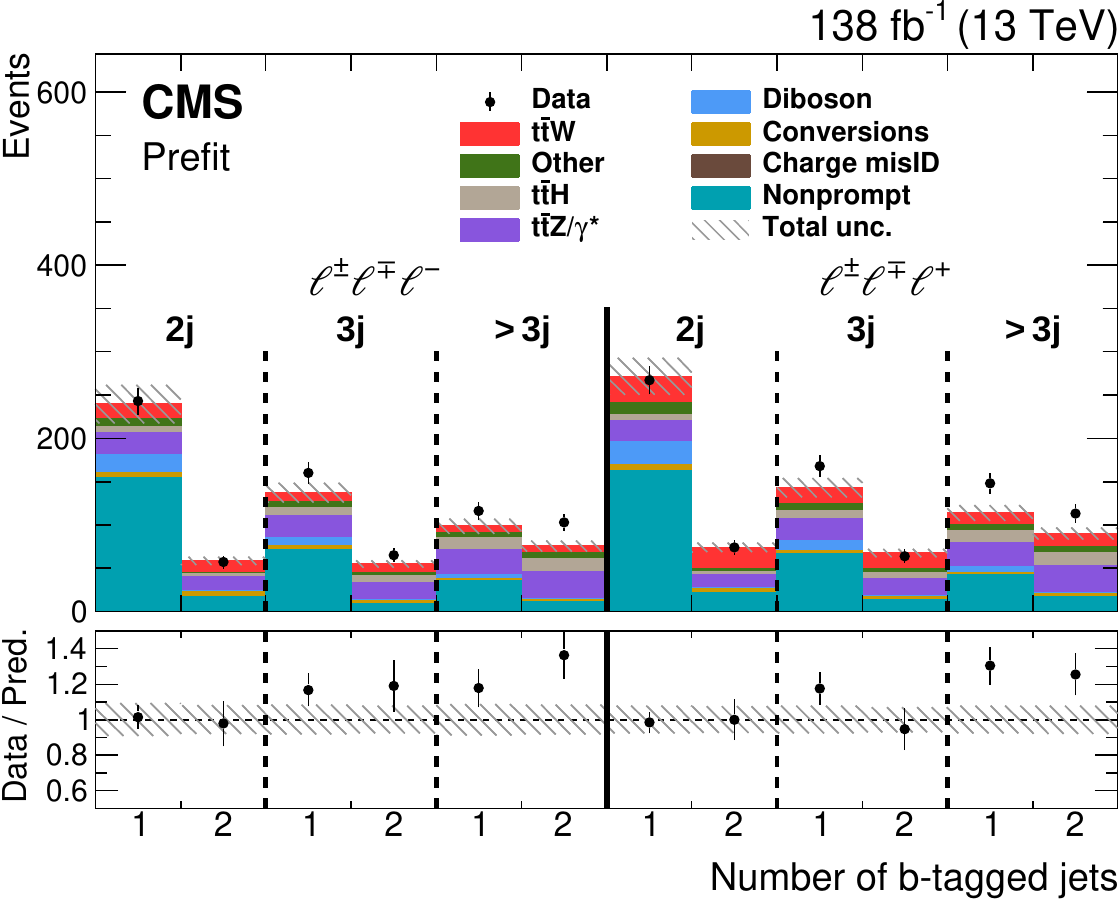}
\caption{%
    Distribution of the number of medium \PQb-tagged jets per event from data
    (points) and prediction (colored histograms) for events in the trilepton
    signal region with different numbers of jets (\PQj).
    Events with total trilepton charge $-1$ and $+1$ are shown in the left and
    right halves of the plot, respectively.
    The predictions are shown before the fit to data (``Prefit'').
    The vertical bars on the points represent the statistical uncertainties in
    the data, and the hatched bands the systematic uncertainty in the
    predictions.
    The lower panel shows the ratio of the event yields in data to the overall
    sum of the predictions.
}
\label{fig:sel3l}
\end{figure}

The analysis strategy in the trilepton channel relies on an event categorization
based on the number of jets and medium \PQb-tagged jets, and on the charges of
the selected leptons.
The number of observed and predicted events in all twelve categories used in the
fit is shown in Fig.~\ref{fig:sel3l}.
The invariant mass of the three leptons, \mlll, was identified as the observable
with the best discriminating power between signal and background contributions,
and is used across all event categories as the discriminant variable.

\section{Background estimation}
\label{sec:backgrounds}

The nonprompt-lepton background contributes to both the same-sign dilepton and
trilepton channels.
Its contribution is estimated from control samples in data using a
tight-to-loose ratio method~\cite{CMS:SUS-15-008, CMS:HIG-17-018,
CMS:HIG-19-008}.
The lepton misidentification rate is defined as the probability \ttl for a loose
lepton to also satisfy the tight selection.
It is parameterized as a function of the lepton \pt and \abseta, and measured in
a data sample enriched in QCD multijet events (``multijet sample'').
The nonprompt-lepton background contribution in the signal channels is then
estimated by applying the measured misidentification rate as a weight to events
that pass the event selection but have at least one ``loose-not-tight'' lepton,
\ie, a lepton that passes the loose but fails the tight selection
(``loose-lepton sample'').
The weights \evw are defined as
\begin{linenomath}\begin{equation}
    \evw=(-1)^{\num+1}\prod_{i=1}^\num\frac{\ttl_i}{1-\ttl_i},
\end{equation}\end{linenomath}
where the product extends over all electrons and muons that pass the loose but
fail the tight selection, and \num refers to the total number of such leptons.
The factor of $(-1)^{\num+1}$ ensures that contributions from events with different numbers of nonprompt leptons are properly taken into account~\cite{CMS:HIG-17-018}.
In both the multijet and loose-lepton sample, the prompt-lepton contributions
are estimated from simulated event samples and subtracted.

Events in the multijet sample are required to have exactly one loose lepton, at
least one jet that is separated from the lepton by $\DR>0.7$, and
$\ptmiss<20\GeV$ to reduce contributions from prompt \Wjets events.
The numbers of multijet events with a loose-not-tight and a tight lepton are
individually determined for each lepton \pt and \abseta bin from maximum
likelihood fits to the transverse mass variable
\begin{linenomath}\begin{equation}
    \mtfix=\sqrt{2\ptfix\,\ptmiss\,[1-\cos(\DPhi)]},
\end{equation}\end{linenomath}
where $\ptfix=35\GeV$ is a constant, and \DPhi is the azimuthal angle between
the lepton momentum and \ptvecmiss.
This \mtfix definition exploits the fact that the \ptmiss reconstructed in QCD
multijet events is mainly due to resolution effects and is typically small,
resulting in a falling distribution of \mtfix.
In contrast, the \mtfix distribution for \Wjets events, which is the main
source of prompt-lepton backgrounds in the QCD multijet selection, has a broad
maximum around 80\GeV.
Since we use \ptfix instead of the reconstructed lepton \pt, the value of \mtfix
does not depend on the latter, allowing for a good separation of QCD multijet
and \Wjets events over the full \pt range~\cite{CMS:HIG-19-008}.
In each \pt and \abseta bin, the lepton misidentification rate is then measured
as $\ttl_i=\Npass/(\Npass+\Nfail)$, where \Npass and \Nfail are the numbers of
events where the lepton is loose-not-tight or tight, respectively, as extracted
from the fit and after subtracting the fitted contributions from prompt leptons.

\begin{figure}[htb!]
\centering
\includegraphics[width=0.42\textwidth]{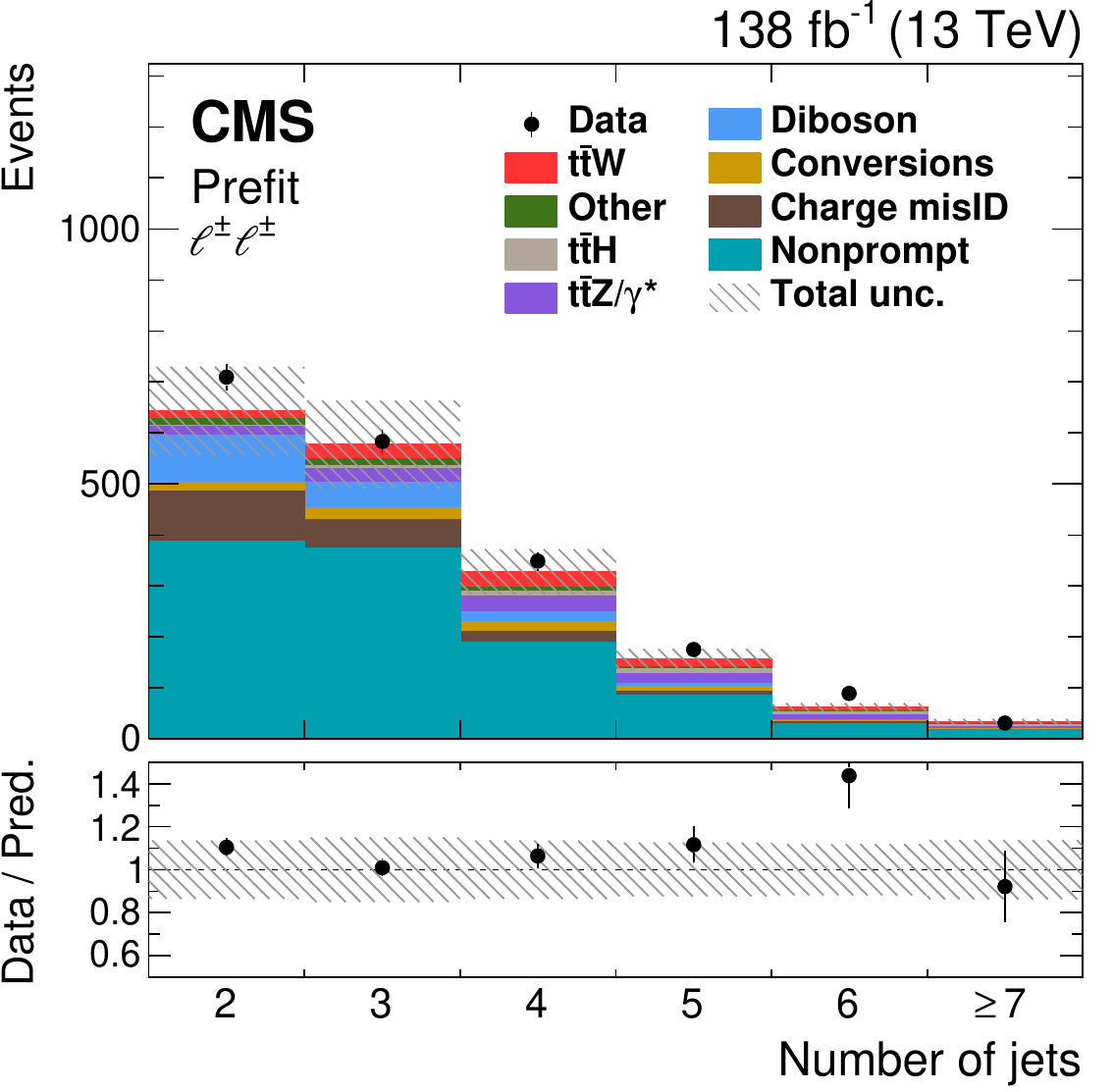}%
\hspace{0.05\textwidth}%
\includegraphics[width=0.42\textwidth]{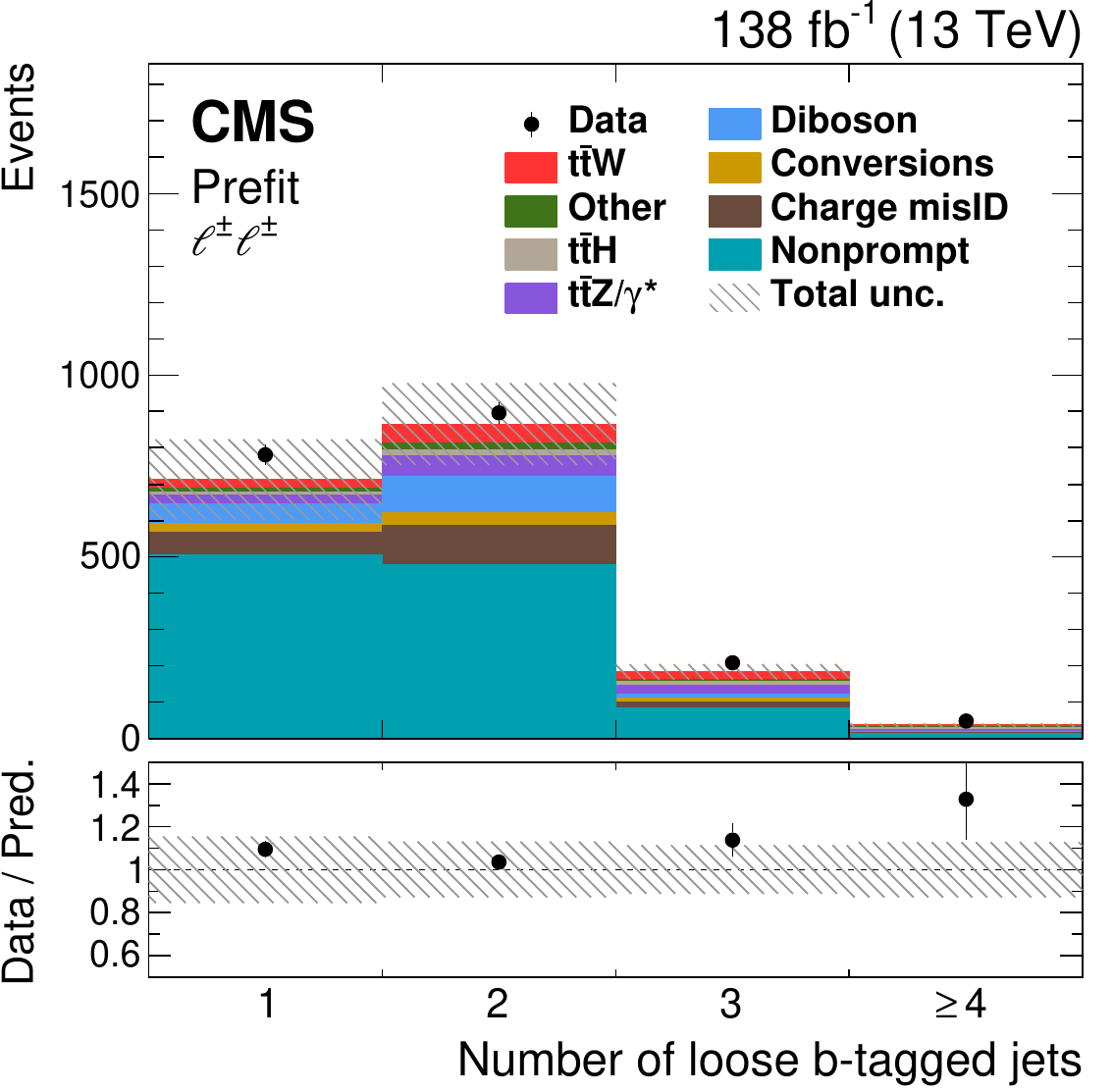} \\[1ex]
\includegraphics[width=0.42\textwidth]{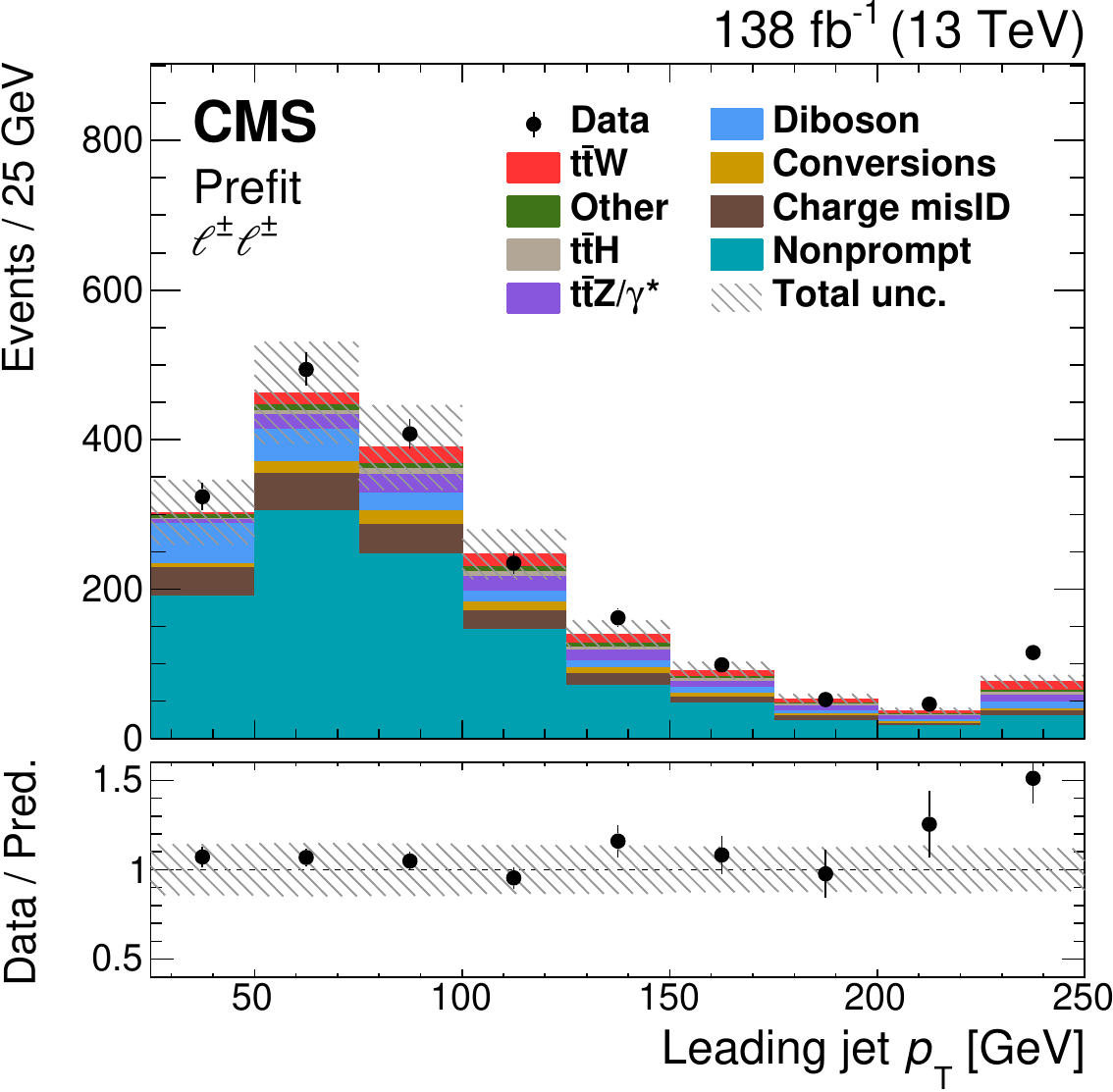}%
\hspace{0.05\textwidth}%
\includegraphics[width=0.42\textwidth]{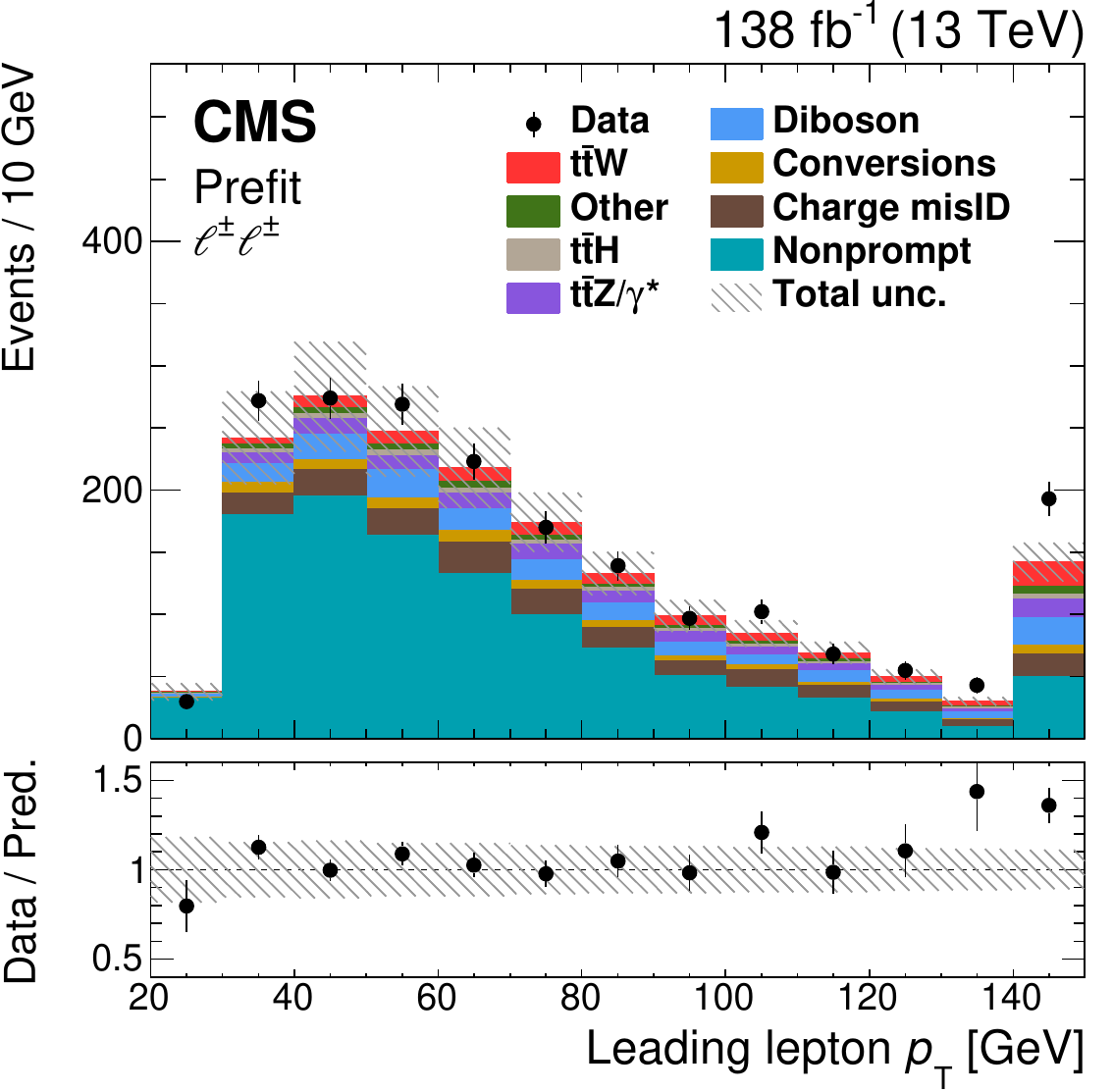}
\caption{%
    Comparison of the number of observed (points) and predicted (colored
    histograms) events in the validation region for nonprompt leptons.
    The distributions of the number of jets (upper left) and loose \PQb-tagged
    jets (upper right), and of the \pt of the leading jet (lower left) and of
    the leading lepton (lower right) are displayed.
    The predictions are shown before the fit to data (``Prefit'').
    The vertical bars on the points represent the statistical uncertainties in
    the data, and the hatched bands the systematic uncertainty in the
    predictions.
    The last bins include the overflow contributions.
    In the lower panels, the ratio of the event yields in data to the overall
    sum of  the predictions is presented.
}
\label{fig:nonprompt}
\end{figure}

The nonprompt-lepton background prediction with the tight-to-loose ratio method
is validated using same-sign dilepton events from data satisfying the full event
selection but with a $\ptmiss<30\GeV$ requirement, making this validation region
orthogonal to the signal region.
The comparison of the distributions from data and prediction of four NN input
variables in the validation region, shown in Fig.~\ref{fig:nonprompt}, shows
good agreement.
This is also true for the other NN input variables, demonstrating that the
prediction provides a good description of the kinematic variables used in the NN
training for the nonprompt-lepton background.
This validation region is used to assign an additional systematic uncertainty in
the nonprompt-lepton background prediction, as described in
Section~\ref{sec:systematics}.

Background contributions arising from the mismeasured electron charge (indicated
as ``Charge misID'' in the figures) are relevant in the same-sign dilepton
channel.
The mismeasurement of the electron charge is typically the result of the
emission of a hard bremsstrahlung photon that undergoes asymmetric conversion.
The reconstructed lepton is then typically the electron or positron that
receives most of the energy of the converted photon, and thus has an equal
probability to have either the same or opposite charge compared to the original
electron or positron~\cite{CMS:EGM-17-001}.
The charge misidentification rate is defined as the probability that the
reconstructed charge of an electron differs from its true value.
It is parameterized as a function of the lepton \pt and \abseta, and is measured
in simulated samples of DY and \ttbar production using tight electrons.
The predicted number of background events is estimated from data by applying the
misidentification rates as weights to events that pass the dilepton event
selection but have two opposite-sign electrons.
The charge misidentification rate for muons was found to be negligible.

Background processes with one or more reconstructed electrons due to the
conversion of a photon are estimated from simulated event samples.
In these events, an electron or a positron typically carries most of the energy
of the converted photon, while the other lepton is of low energy and not
reconstructed.
The largest source of events with electrons from photon conversions in the
signal regions is from \ttg production.

Background contributions with prompt leptons are estimated from simulated event
samples, and are normalized to the cross sections predicted from SM
calculations.
The main contributions arise from \ttH, \ttZG, and diboson (\WZ and \ZZ)
production.
Smaller contributions, including \VVV and rare top quark (\ttVV, \tttt, \tHq,
and \tZq) production, are grouped together and are indicated as ``Other'' in the
figures.

\begin{figure}[htp!]
\centering
\includegraphics[width=0.45\textwidth]{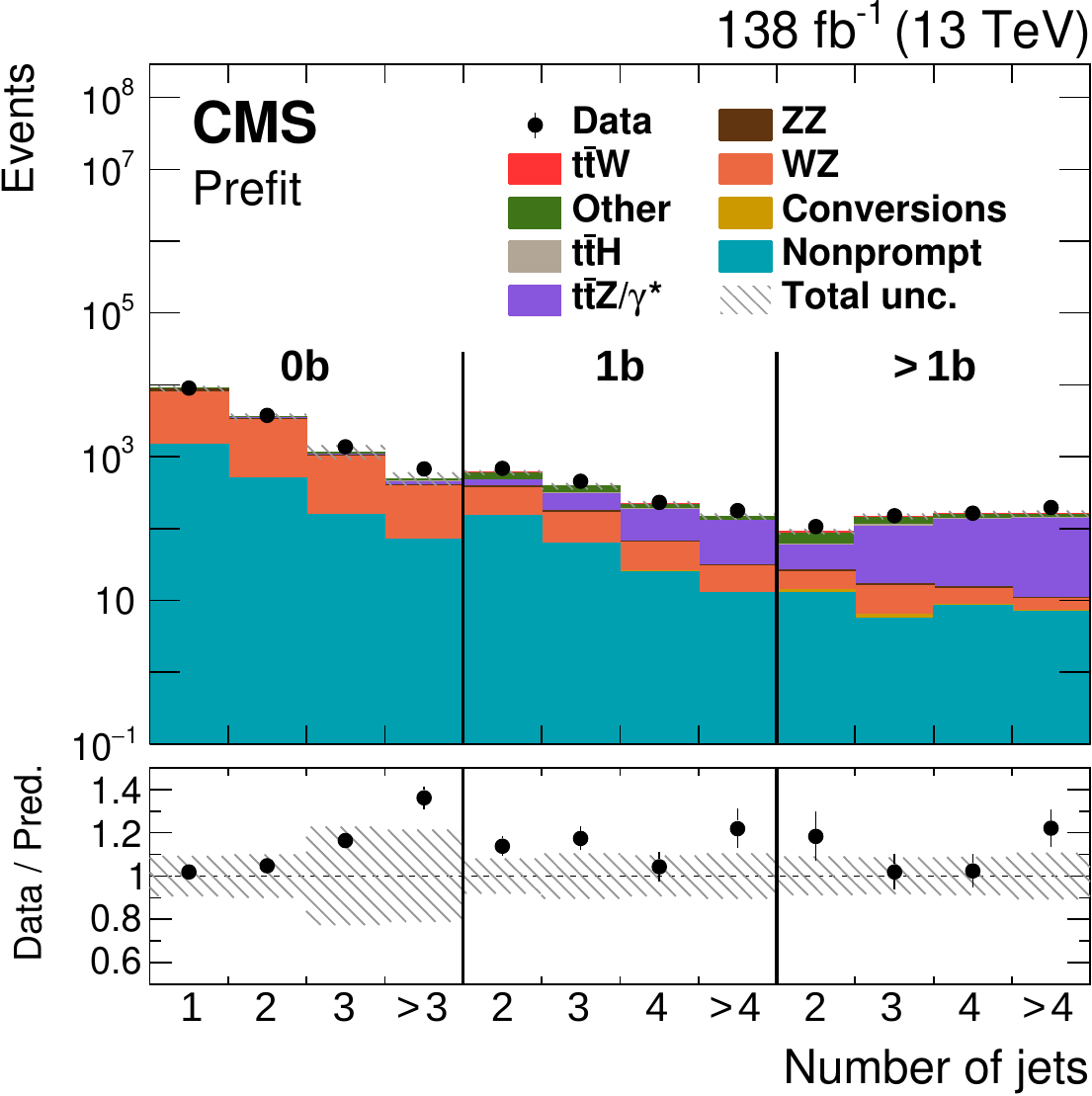}
\caption{%
    The distributions of the number of jets in events from the control region
    enriched in \WZ and \ttZ production from data (points) and prediction
    (colored histograms).
    The events are divided into three groups depending on the number of medium
    \PQb-tagged jets \PQb in the event.
    The predictions are shown before the fit to data (``Prefit'').
    The vertical bars on the points represent the statistical uncertainties in
    the data, and the hatched bands the systematic uncertainty in the
    predictions.
    In the lower panels, the ratio of the event yields in data to the overall
    sum of the predictions is presented.
}
\label{fig:cr3l}
\end{figure}

For the trilepton channel, the predictions from simulated event samples for the
dominant processes with a \PZ boson (\WZ, \ZZ, and \ttZ) are validated using
dedicated data control regions.
Events are selected for a \WZ and \ttZ production control region by applying the
full event selection except for the requirement on same-flavor opposite-sign
lepton pairs, where we require that there be at least one pair with
$\DmllZ<10\GeV$.
This ensures that the control region is orthogonal to the trilepton signal
region.
The events are grouped according to the number of jets and medium \PQb-tagged
jets, and the distributions of these quantities from data and prediction are
compared in Fig.~\ref{fig:cr3l}.
The control region is further split into four categories based on the flavors of
the selected leptons, which are included in the fit for the extraction of the
\ttW cross section.

\begin{figure}[htp!]
\centering
\includegraphics[width=0.45\textwidth]{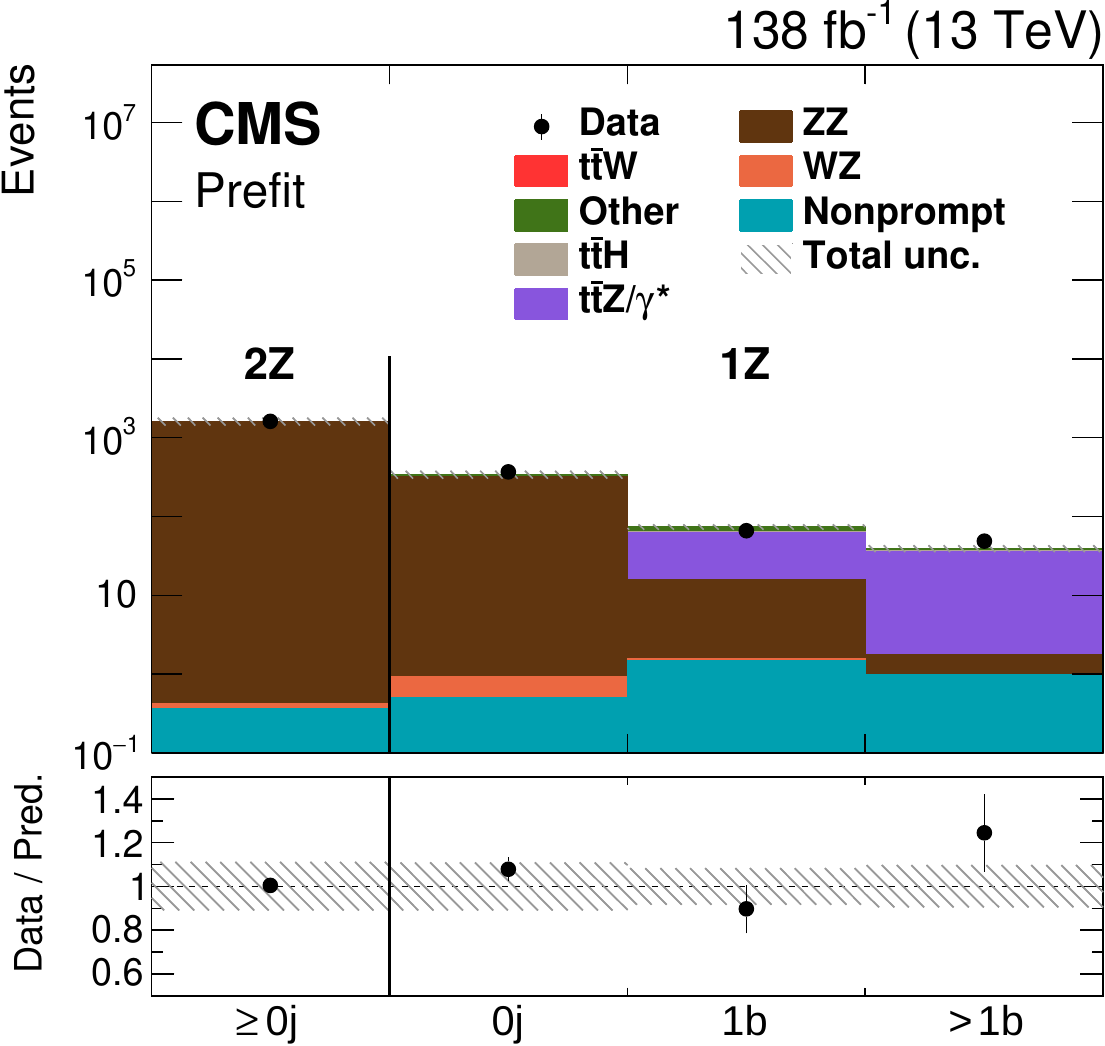}
\caption{%
    Distribution of the number of jets \PQj and medium \PQb-tagged jets \PQb in
    events from the control region enriched in \ZZ and \ttZ events for data
    (points) and prediction (colored histograms).
    Events in the first bin have two \PZ boson candidates (2\PZ) and events in
    the other bins have exactly one \PZ boson candidate (1\PZ).
    The predictions are shown before the fit to data (``Prefit'').
    The small contribution from background processes with nonprompt leptons is
    estimated from simulated event samples.
    The vertical bars on the points represent the statistical uncertainties in
    the data, and the hatched bands the systematic uncertainty in the
    predictions.
    In the lower panels, the ratio of the event yields in data to the overall
    sum of the predictions is presented.
}
\label{fig:cr4l}
\end{figure}

A control region enriched in \ZZ and \ttZ events is defined by an event
selection that requires exactly four tight leptons.
Ordering the leptons in decreasing order of momentum, they are required to have
$\pt>25$, 15, 15, and 10\GeV, respectively.
Additionally, the presence of at least one \PZ boson candidate, defined as a
same-flavor opposite-sign lepton pair with $\DmllZ<10\GeV$, is required.
The remaining events are then categorized into four groups: (1) events with two
\PZ boson candidates and any number of jets; (2) events with no second \PZ boson
candidate and no jets; (3) events with no second \PZ boson candidate and exactly
one medium \PQb-tagged jet; and (4) events with no second \PZ boson candidate
and at least two medium \PQb-tagged jets.
Events in categories (3) and (4) may have any number of additional jets that are
not medium \PQb tagged.
Events not passing any of these criteria are removed from the control region.
The distributions of the number of jets and medium \PQb-tagged jets for this
control region are shown in Fig.~\ref{fig:cr4l} from data and prediction.
The four groups are also included in the fit for the extraction of the \ttW
cross section, in order to constrain the normalization of the \ZZ and \ttZ
backgrounds.

\section{Systematic uncertainties}
\label{sec:systematics}

Multiple sources of systematic uncertainty affect the signal and background
event yields and the distributions of the discriminating observables used for
the signal extraction.
These sources can be divided into two categories.
Sources that affect only the yield of the signal or background processes are
considered as rate uncertainties.
Sources that also modify the shape of the fitted distributions are referred to
as shape uncertainties.
The main sources of experimental and theoretical systematic uncertainties are
described below.

\textbf{Integrated luminosity:}\quad
The integrated luminosities for the 2016, 2017, and 2018 data-taking years have
individual uncertainties between 1.2 and 2.5\%~\cite{CMS:LUM-17-003,
CMS:LUM-17-004, CMS:LUM-18-002}, while the overall uncertainty for the
2016--2018 period is 1.6\%.
The uncertainty in the integrated luminosity is a rate uncertainty and affects
both the normalization of the background contributions from simulated event
samples, as well as the extraction of the measured cross section from the final
estimate of the number of signal events.

\textbf{Pileup reweighting:}\quad
The distribution of the number of additional \pp interactions per event in
simulation is matched to data by reweighting the profile of the true number of
interactions to the one inferred from the instantaneous luminosity profile in
data.
A systematic uncertainty is estimated by varying the assumed minimum-bias cross
section of 69.2\unit{mb} by $\pm$4.6\%~\cite{CMS:FWD-11-002, CMS:FWD-11-001}.
This variation is propagated throughout the analysis, yielding a shape
uncertainty that is considered fully correlated among the data-taking years.

\textbf{Trigger efficiency:}\quad
The efficiency of the trigger selection is measured in data and simulated events
using independent trigger paths based on hadronic activity or \ptmiss
signatures.
Per-event weights that depend on the lepton \pt are applied to the simulated
samples in order to correct for discrepancies.
The impact on the final discriminants due to the trigger efficiency uncertainty
is estimated by varying these scale factors within their uncertainties, which
are determined to be 2\%.
This shape uncertainty is treated as uncorrelated among data-taking years, as
well as between the dilepton and trilepton channels.
A separate uncertainty is determined for each lepton flavor category in the
dilepton channel.

\textbf{L1 inefficiency:}\quad
During the 2016 and 2017 data-taking periods, a gradual shift in the timing of
the inputs of the ECAL L1 trigger in the region at $\abseta>2$ caused a specific
trigger inefficiency.
For events containing an electron (a jet) with \pt larger than $\approx$50\GeV
($\approx$100\GeV), in the region $2.5<\abseta<3.0$ the efficiency loss is
$\approx$10--20\%, depending on \pt, \sigeta, and time.
Correction factors were computed from data and applied to the acceptance
evaluated by simulation, resulting in a shape uncertainty that is correlated
between 2016 and 2017.
As expected in analyses that do not rely on forward objects, the effect on the
results is small (below 1\%).

\textbf{Lepton efficiency:}\quad
The efficiency of the tight-lepton selection is measured in data and simulation
using the ``tag-and-probe'' method in $\PZ\to\Pellp\Pellm$
events~\cite{CMS:MUO-16-001, CMS:EGM-17-001}.
Per-lepton corrections are derived separately for each lepton flavor, and a
shape uncertainty is obtained and found to be at most a few percent in each bin.
This is treated as correlated between data-taking years, and uncorrelated
between the lepton flavors.

\textbf{Jet energy scale and resolution:}\quad
The variations of the jet momenta due to the uncertainty in the jet energy scale
and resolution are composed of 21 uncertainty sources which account for
different detector regions and the year-to-year
correlations~\cite{CMS:JME-13-004}.
The uncertainties in the \ptmiss resolution and response are accounted for by
varying the jet energy scale in simulation within their respective uncertainties
and recomputing \ptmiss after each variation.
The \ptmiss uncertainties are not relevant for the  trilepton channel, where
\ptmiss is not used.
This resulting shape uncertainty is considered to be partially correlated
between data-taking years.

\textbf{\PQb tagging efficiency:}\quad
To match the \PQb tagging efficiency in simulation to data, the full
DeepJet discriminant shape is corrected with scale factors measured with the
``iterative fit'' method~\cite{CMS:HIG-13-029, CMS:BTV-16-002}.
The uncertainty in the \PQb tagging efficiency scale factors is considered to be
fully correlated between \PQb and \PQc quark jets, and uncorrelated for other
quark flavors.
It is applied as a function of \pt and \abseta of the jets, and propagated as
shape uncertainty to the fitted distributions.
These sources are considered as partially correlated between data-taking years.

\textbf{Background normalization:}\quad
Several rate uncertainties are considered for the normalization of the
background processes.
While the initial normalizations of the \ttZ, \WZ, and \ZZ background
contributions are taken from SM predictions, the inclusion of the control
regions described in Section~\ref{sec:backgrounds} provides strong constraints
on these values.
The \WZ and \ZZ events with a large number of \PQb-tagged jets are either due to
misidentified light-flavor jets, or to the inclusion of a phase space region
that is not well described in simulation.
This has been studied in Ref.~\cite{CMS:HIG-17-018}, and we apply an additional
40 (10)\% uncertainty to the predicted number of events with two or more (fewer
than two) \PQb-tagged jets, following those results.
An additional uncertainty of 30\% is applied to \WZ and \ZZ events with at least
three selected jets in order to account for mismodeling effects observed in the
corresponding control region described in Section~\ref{sec:backgrounds}.
The normalization of the \ttH background is constrained within a 20\%
uncertainty~\cite{CMS:HIG-19-008}.
The normalization of the \ttg process includes an uncertainty of
8\%~\cite{CMS:TOP-18-010}.
To account for a possible mismodeling of photon conversions in the simulation,
an additional 30\% uncertainty is applied to the predicted number of events,
following the studies performed in Ref.~\cite{CMS:SUS-16-039}.
The normalizations of the \tHq and \tHW backgrounds are each constrained within
a 50\% uncertainty~\cite{CMS:HIG-22-001}.
The \tZq normalization uncertainty amounts to 10\%, following the latest
measurement done by CMS~\cite{CMS:TOP-20-010}.
A 50\% uncertainty is applied to the \ttVV normalization, which reflects the
current experimental precision~\cite{CMS:TOP-17-009}.
A 50\% uncertainty is estimated for the triboson production, as the main
contribution to this background comes from the $\PW\PW\PW$ production, studied
in Ref.~\cite{CMS:SMP-19-014}.
For other background processes with smaller production cross sections, the
yields are constrained within 50\%.
Even if several of these processes have been measured to a higher precision, the
assigned uncertainty conservatively takes into account the extrapolation of
these predictions to the kinematic region important for this analysis, which
contains events with high jet and \PQb-tagged jet multiplicities.

\textbf{Nonprompt-lepton background:}\quad
The misidentification rates used to estimate the nonprompt-lepton background are
affected by a statistical uncertainty associated with the multijet sample, as
well as inaccuracies in the modeling of the contamination due to prompt leptons
in this sample.
For each lepton flavor, shape uncertainties are added to the prediction of the
background processes with nonprompt leptons.
These account for the overall normalization, as well as for the \pt and \sigeta
dependence of the statistical uncertainty and the observed mismodeling effects.
These are taken as correlated among the data-taking years.
An additional systematic uncertainty of 20\% is applied to the
nonprompt-background estimate that covers the differences between the prediction
and the observation in the validation region for nonprompt leptons.
This systematic uncertainty is taken as correlated among the data-taking years
and the considered final states.

\textbf{Charge misidentification background:}\quad
We compare data and simulated events in a selection with exactly two tight
electrons of same sign and $\abs{m(\Pe\Pe)-\mZ}<15\GeV$.
Based on this comparison, we assign a rate uncertainty of 20\% to the charge
misidentification background.
Since this background is negligible for the trilepton channel, this
uncertainty is not included there.

\textbf{Modeling uncertainties:}\quad
Uncertainties in the modeling of the signal and background processes may impact
the acceptance and selection efficiencies, the shape of the fitted
distributions, and the predicted cross sections used for the normalization of
the simulated event samples.
All modeling uncertainties are considered as correlated among data-taking
years.

Uncertainties in the simulated event yields due to missing higher-order diagrams
in the ME calculations are evaluated by varying the values of \muR and \muF.
We vary the scales by a factor of two up and down, separately for each scale as
well as simultaneously by the same factor, and evaluate the envelope of the six
resulting variations to determine the uncertainty in each bin.
The resulting shape uncertainty is evaluated for all backgrounds that are
estimated from simulation.
For the signal process, the effect from these variations on the predicted cross
section is not included as a systematic uncertainty.
We find that the effect of an additional uncertainty in the relative weight of
the signal samples corresponding to the EWK and QCD terms is negligible.

For the signal process, the effect of the energy scale in initial- and
final-state radiation (ISR and FSR) in the PS simulation is taken into account
as a shape uncertainty.
The effect of the color-reconnection model is evaluated from a comparison of
simulated \ttW events simulated with alternative models~\cite{CMS:GEN-17-002},
and an uncertainty of 1\% is assigned to cover the observed differences.
Comparisons to higher-order generators, which have been performed \eg in 
Refs.~\cite{FebresCordero:2021kcc, Ferencz:2023fso} and could in principle
result in signal modeling deviations not covered by the ME and PS uncertainties
described here, have not been considered.
The effect on the shape of the signal due to the PDF uncertainty is also taken
into account as a shape uncertainty, following the procedure described in
Ref.~\cite{Butterworth:2015oua}.
Using the Hessian NNPDF3.1 set~\cite{EXT:NNPDF-2017}, the difference with
respect to the central value is computed for each of the 100 PDF variations, and
these variations are added in quadrature.
Then, the \alpS variation is taken into account, adding the variation in
quadrature.

\section{Results}
\label{sec:results}

A binned profile likelihood fit is performed on data using the predicted
distributions for the signal and background processes~\cite{CMS:NOTE-2011-005,
EXT:Asymptotic-2011} in order to extract the production rate of the \ttW
process.
The fit is based on a likelihood function $L$ built from the Poisson
probabilities to obtain the observed yields given the predicted signal and
background estimates, and includes terms for the systematic uncertainties and
their correlations.
In the dilepton signal region, the NN output score in eight categories defined
by lepton flavor and charge is used.
In the trilepton signal region, the \mlll distributions are fitted using twelve
event categories defined by the jet and medium \PQb-tagged jet multiplicities,
as well as the charges of the leptons, as defined in
Section~\ref{sec:eventselection}.
The fit further includes the event yields in the control region enriched in \WZ
and \ttZ events (in 48 bins grouped by the number of jets and \PQb-tagged jets,
and lepton flavors), and in the control region enriched in \WZ and \ttZ events
(in four bins grouped by the number of jets, \PQb-tagged jets, and \PZ
candidates), as defined in Section~\ref{sec:backgrounds}.
All of the event categories are further split into the three data-taking years.

Rate uncertainties are represented in the likelihood fit by a log-normal
probability density function, and shape uncertainties are assumed to follow a
Gaussian function.
Statistical uncertainties in the yields per bin are taken into account through
a single nuisance parameter for all
processes~\cite{Barlow:1993dm, Conway:2011in}.

A comparison between the total number of observed and predicted events obtained
before the fit in the dilepton and trilepton channels is shown in
Table~\ref{tab:events}.
Good agreement between data and the total signal plus background prediction
within the systematic uncertainties is observed for each channel.

\begin{table}[htbp!]
\centering
\topcaption{%
    Number of predicted and observed events in the dilepton and trilepton signal
    regions before the fit to the data.
    The uncertainties in the predicted number of events include both the
    statistical and systematic components.
    The uncertainties in the total number of predicted background and background
    plus signal events in each channel are also given.
    The symbol ``\NA'' indicates that the corresponding background does not
    apply.
}
\renewcommand\arraystretch{1.2}
\begin{tabular}{ccccc}
    Process & $\Pellp\Pellp$ & $\Pellm\Pellm$ & $\Pellpm\Pellmp\Pellp$ & $\Pellpm\Pellmp\Pellm$ \\ \hline
    \ttW & $677\pm21$ & $355\pm12$ & $119.4\pm9.2$ & $65.3\pm5.4$ \\[\cmsTabSkip]
    Nonprompt & $2490\pm600$ & $2360\pm570$ & $325\pm75$ & $298\pm71$ \\
    Charge misID & $520\pm110$ & $520\pm111$ & \NA & \NA \\
    \ttH & $167\pm34$ & $169\pm34$ & $56\pm12$ & $57\pm12$ \\
    \ttZG & $335\pm26$ & $333\pm26$ & $145\pm13$ & $147\pm13$ \\
    Diboson & $382\pm88$ & $285\pm65$ & $46.8\pm9.1$ & $38.0\pm7.5$ \\
    Other & $178\pm34$ & $126\pm27$ & $43.4\pm8.2$ & $33.5\pm7.4$ \\
    Conversions & $177\pm54$ & $192\pm59$ & $22.9\pm7.1$ & $24.0\pm7.4$ \\[\cmsTabSkip]
    Total background & $4250\pm620$ & $4000\pm590$ & $639\pm80$ & $600\pm76$ \\
    Total prediction & $4920\pm620$ & $4350\pm590$ & $758\pm81$ & $663\pm76$ \\[\cmsTabSkip]
    Data & 5143 & 4486 & 834 & 744 \\
\end{tabular}
\label{tab:events}
\end{table}

\begin{figure}[htp!]
\centering
\includegraphics[width=0.42\textwidth]{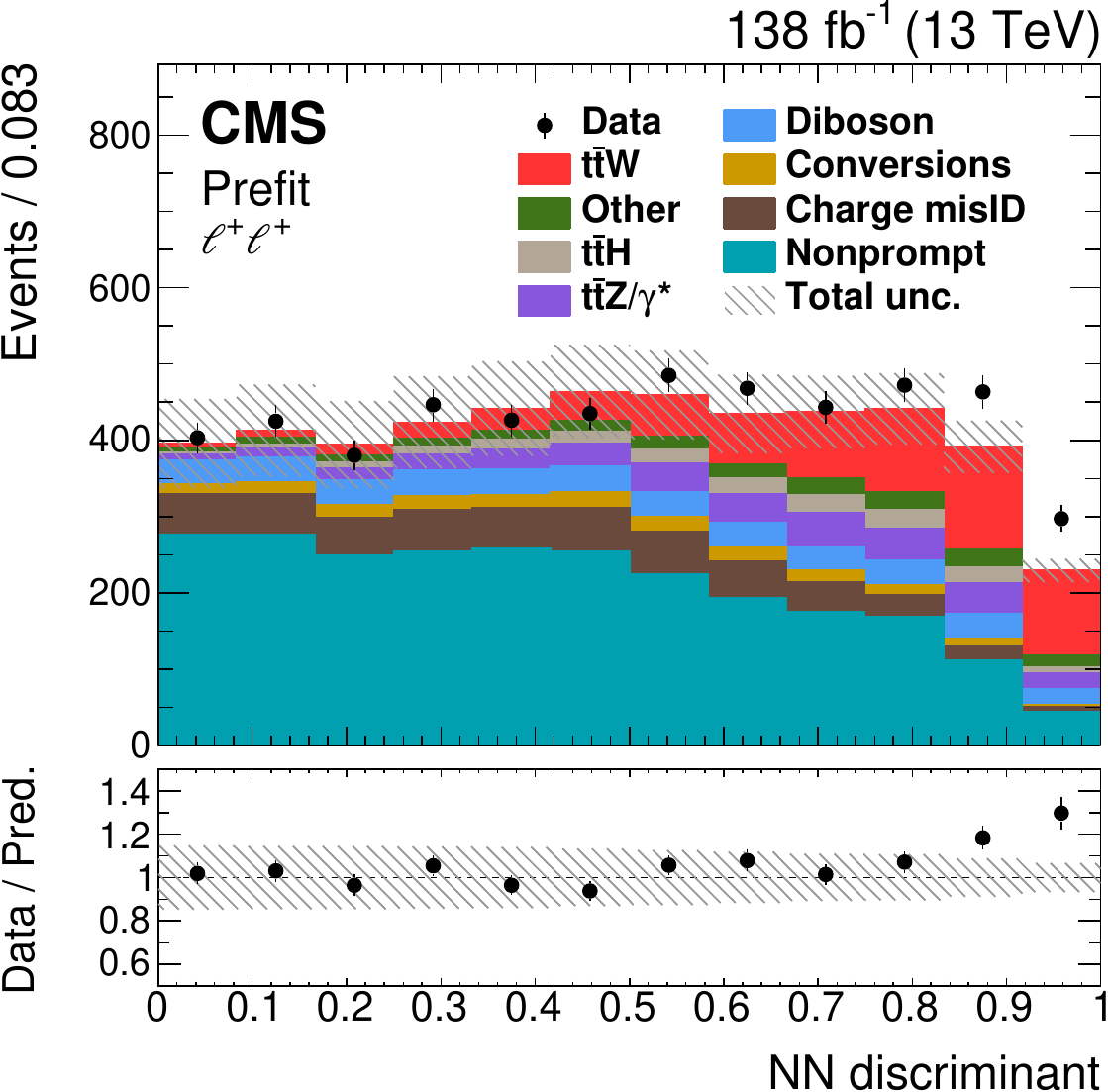}%
\hspace{0.05\textwidth}%
\includegraphics[width=0.42\textwidth]{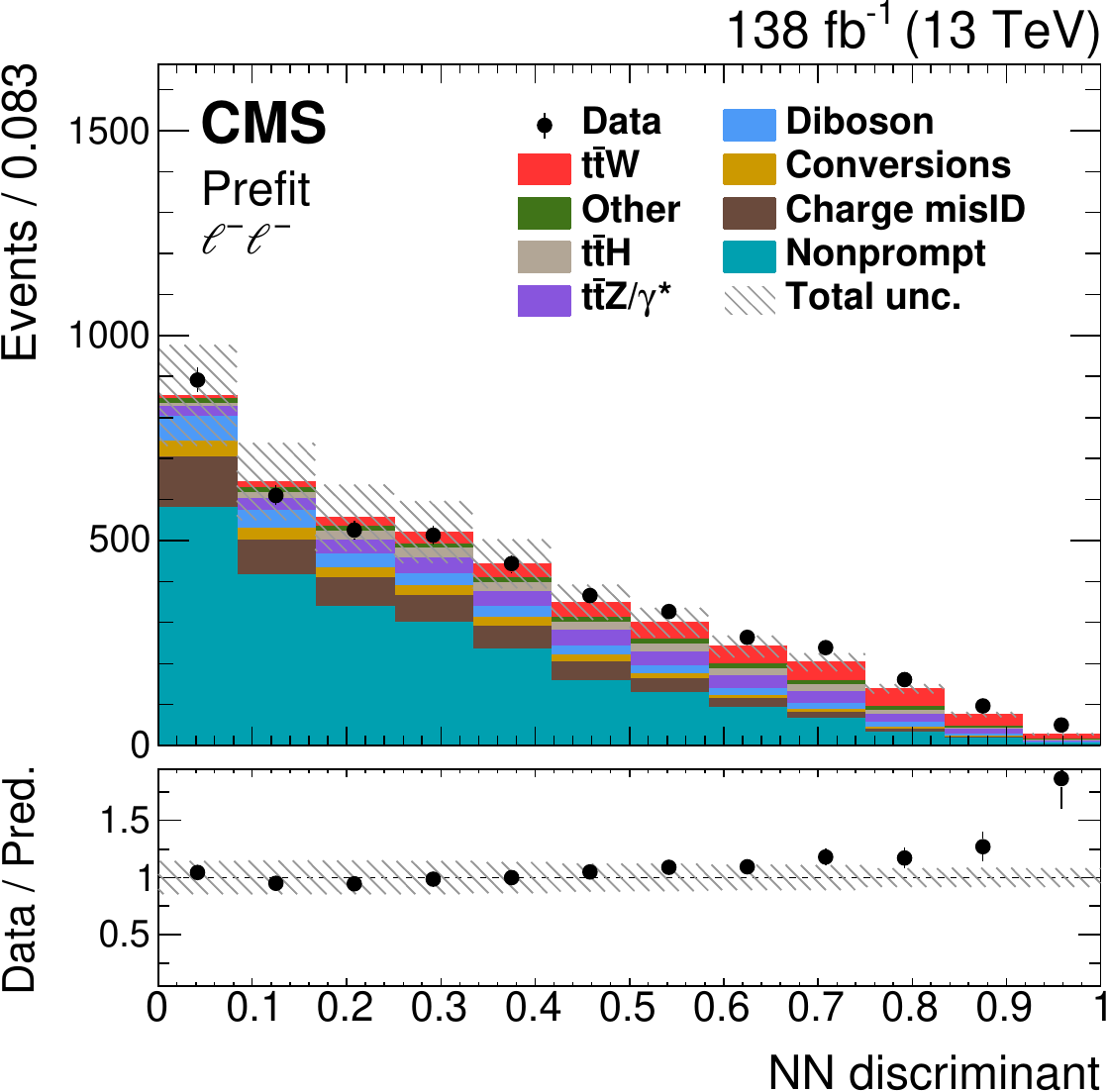} \\[1ex]
\includegraphics[width=0.42\textwidth]{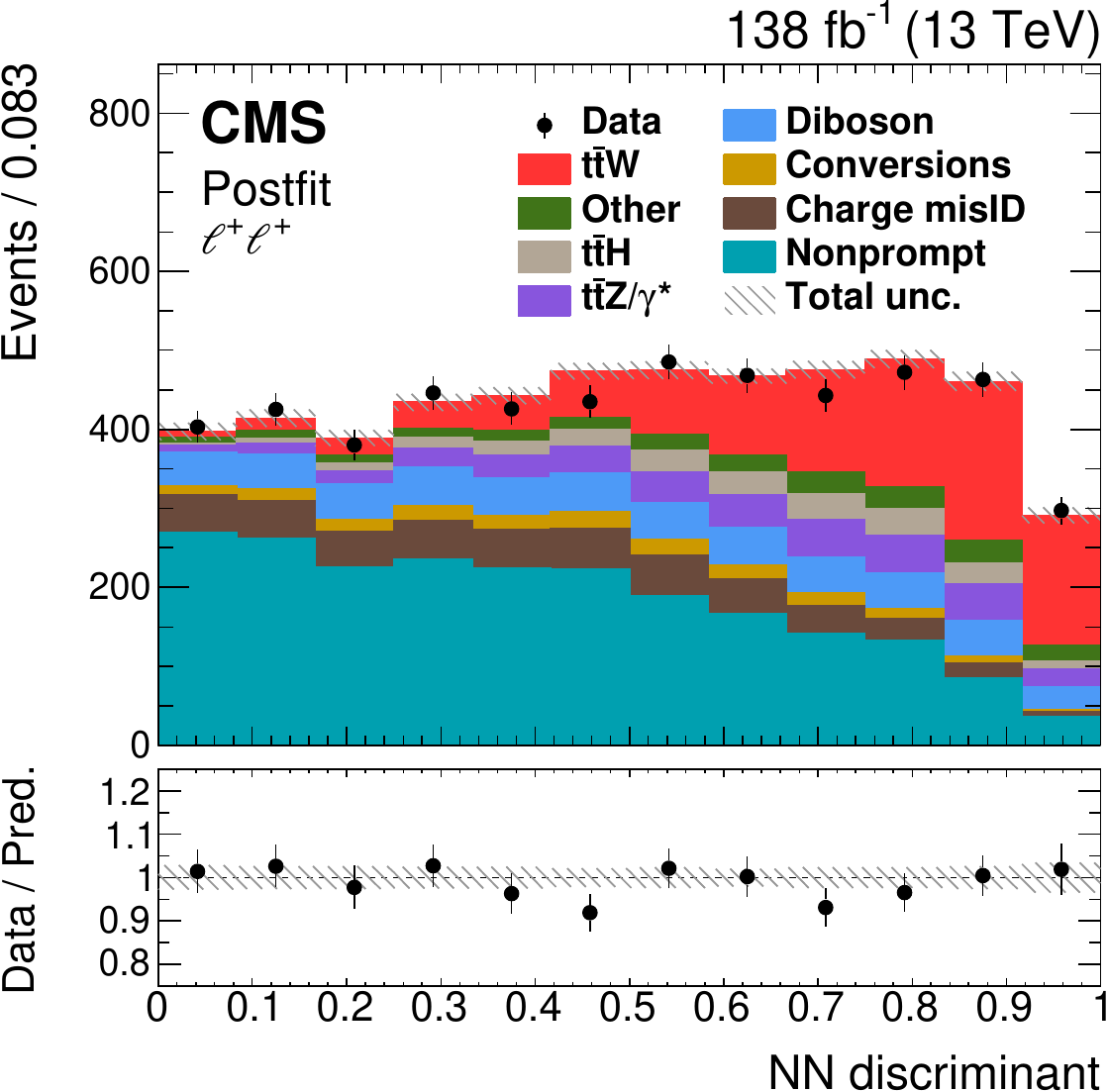}%
\hspace{0.05\textwidth}%
\includegraphics[width=0.42\textwidth]{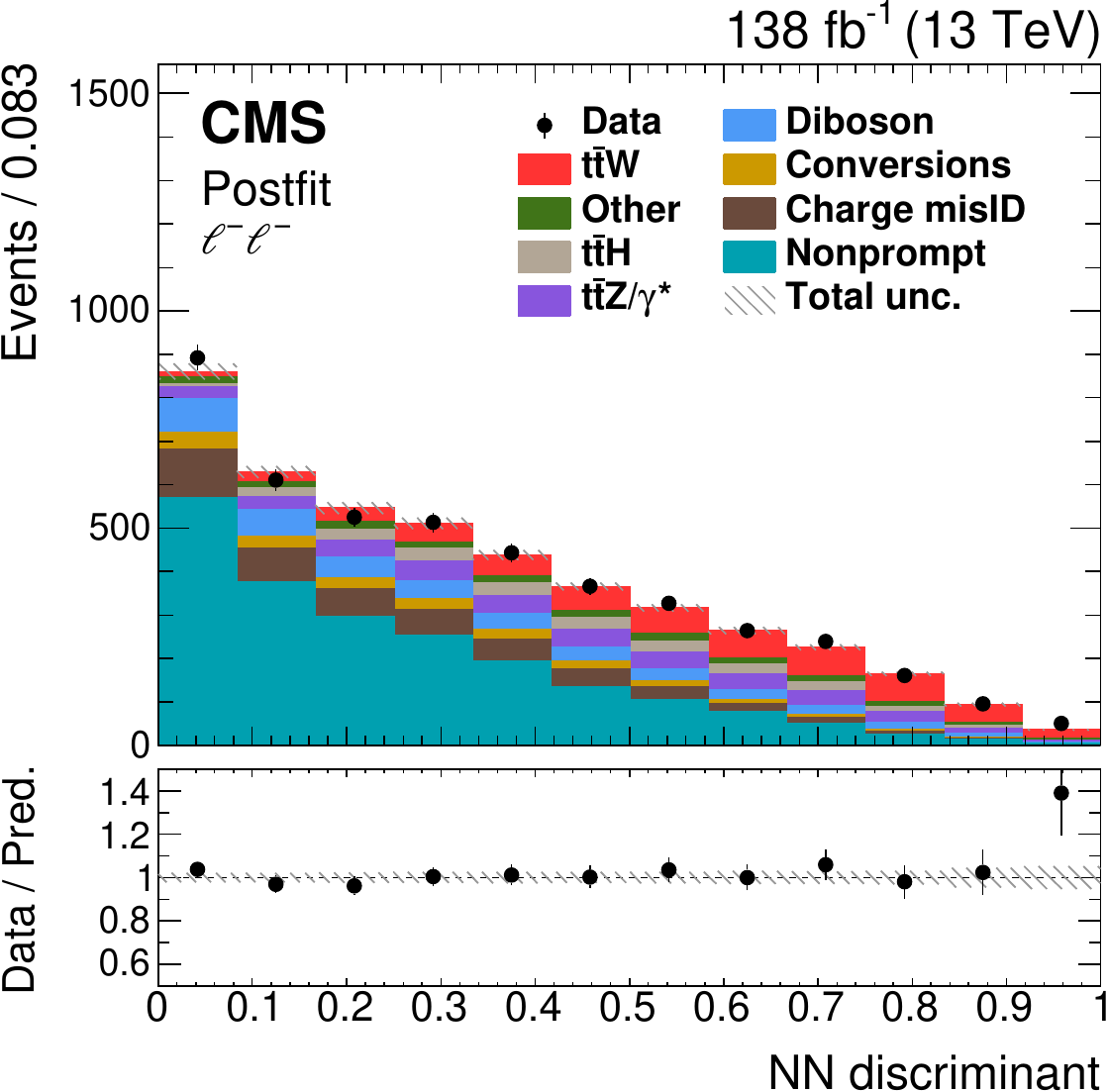}
\caption{%
    Comparison of the number of observed (points) and predicted (colored
    histograms) events in the same-sign dilepton signal region.
    The distributions of the NN output score are displayed in final states with
    two leptons of positive (left) and negative (right) charges.
    The predictions are shown ``prefit'' as in Fig.~\ref{fig:sel2l} (upper row),
    and with the values of the normalizations and nuisance parameters obtained
    in the fit to the data applied (``postfit'', lower row).
    The vertical bars on the points represent the statistical uncertainties in
    the data, and the hatched bands the systematic uncertainty in the
    predictions.
    In the lower panels, the ratio of the event yields in data to the overall
    sum of the predictions is presented.
}
\label{fig:fit2l}
\end{figure}

\begin{figure}[htp!]
\centering
\includegraphics[width=0.42\textwidth]{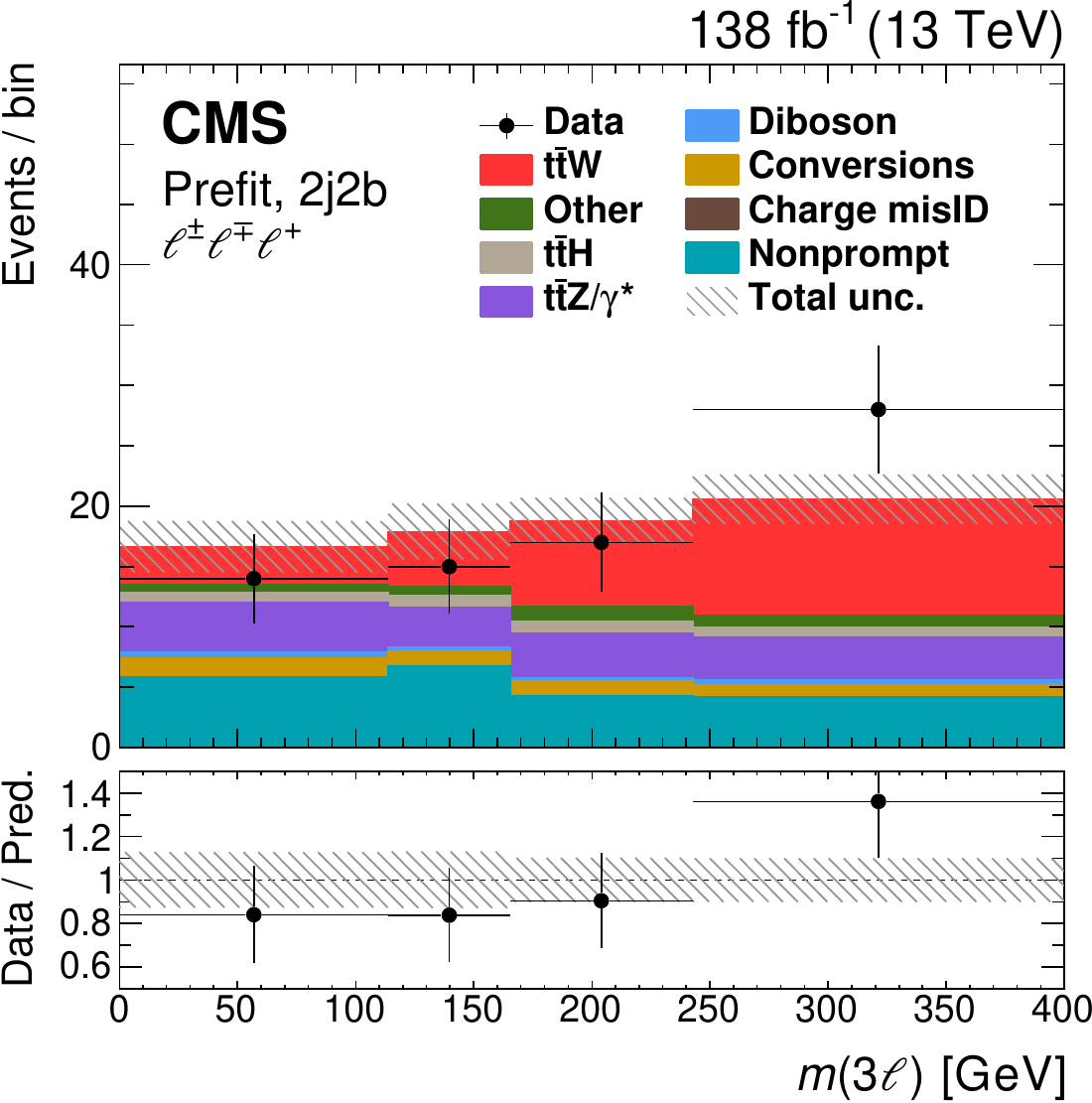}%
\hspace{0.05\textwidth}%
\includegraphics[width=0.42\textwidth]{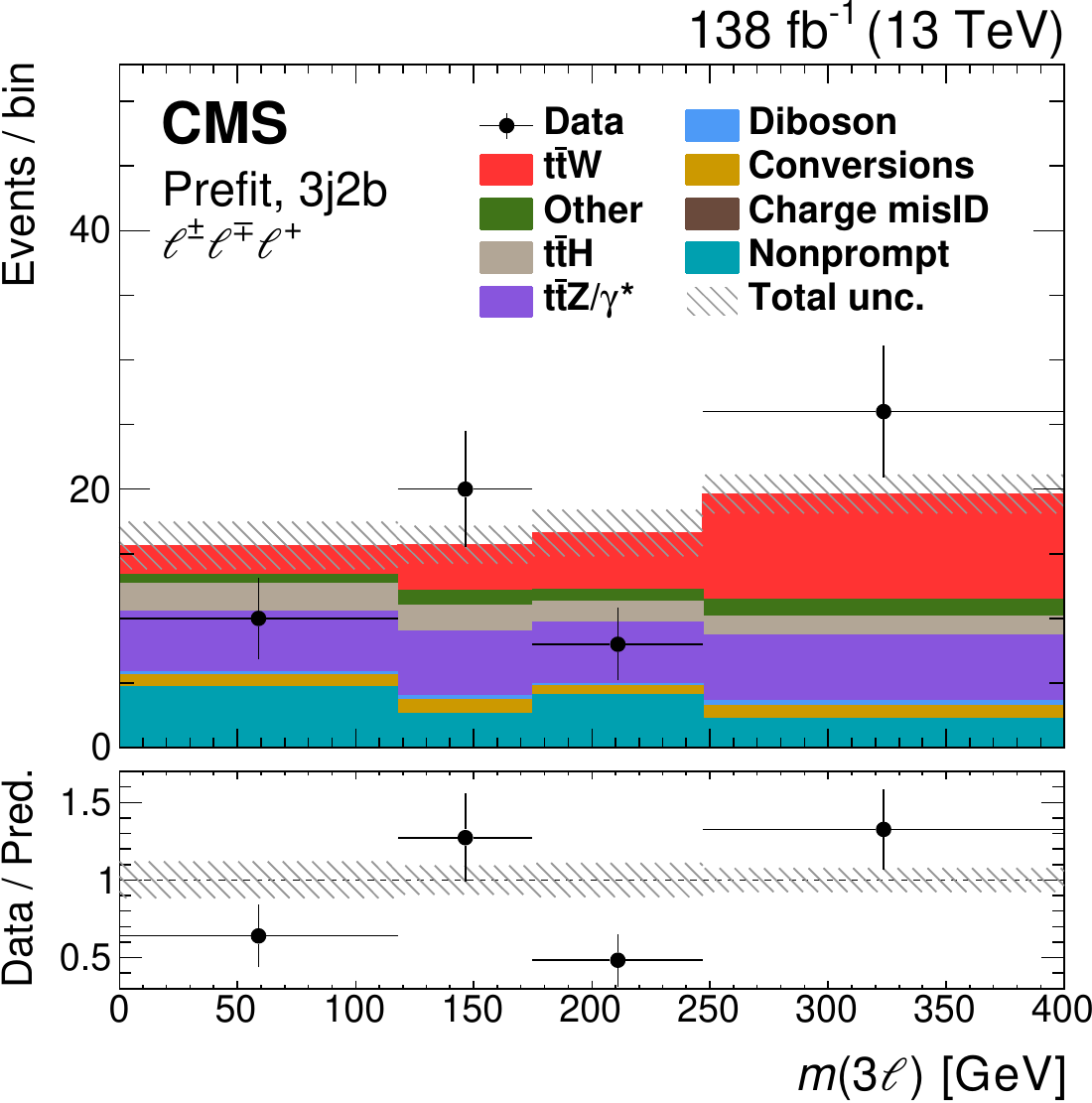} \\[1ex]
\includegraphics[width=0.42\textwidth]{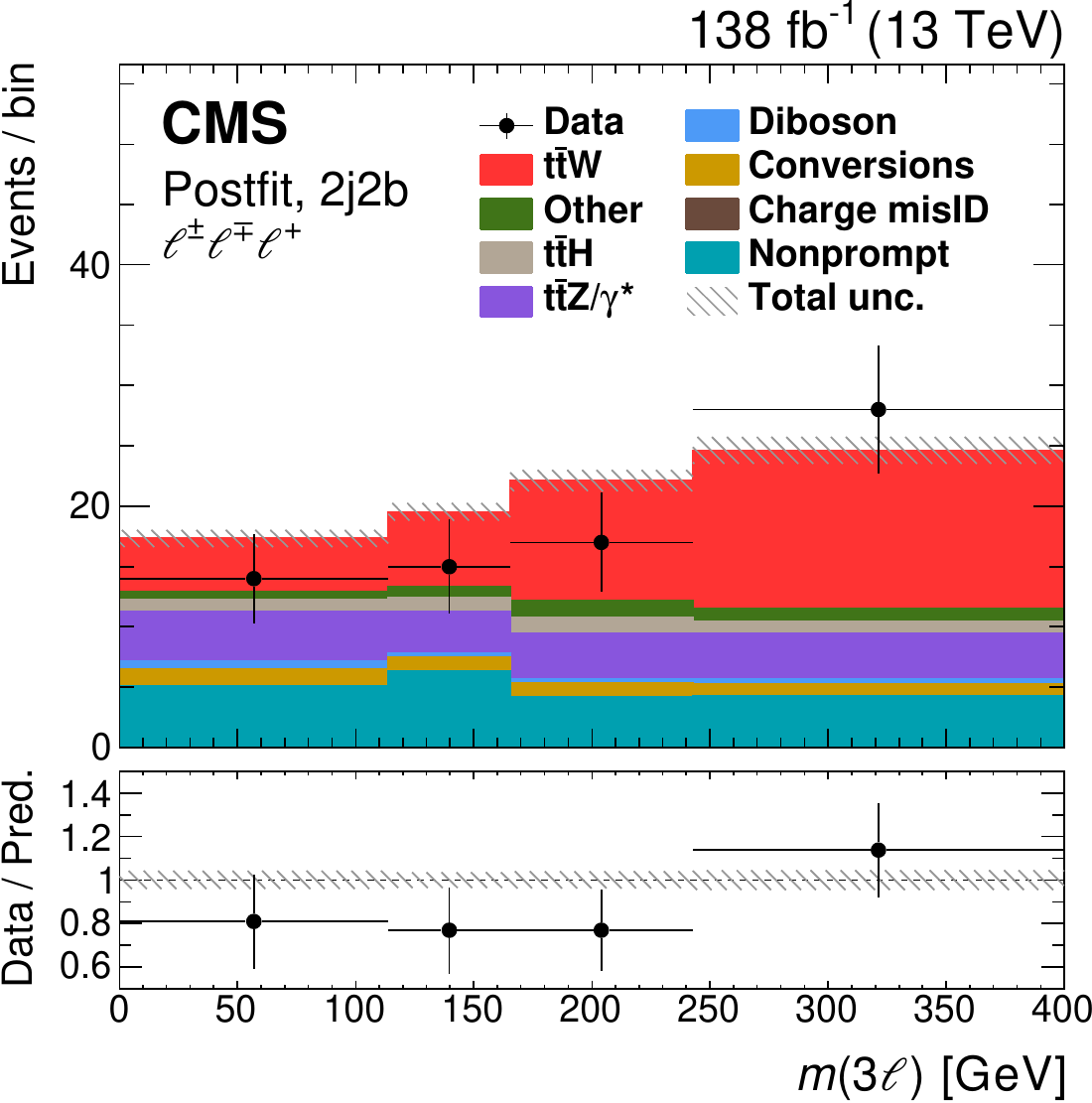}%
\hspace{0.05\textwidth}%
\includegraphics[width=0.42\textwidth]{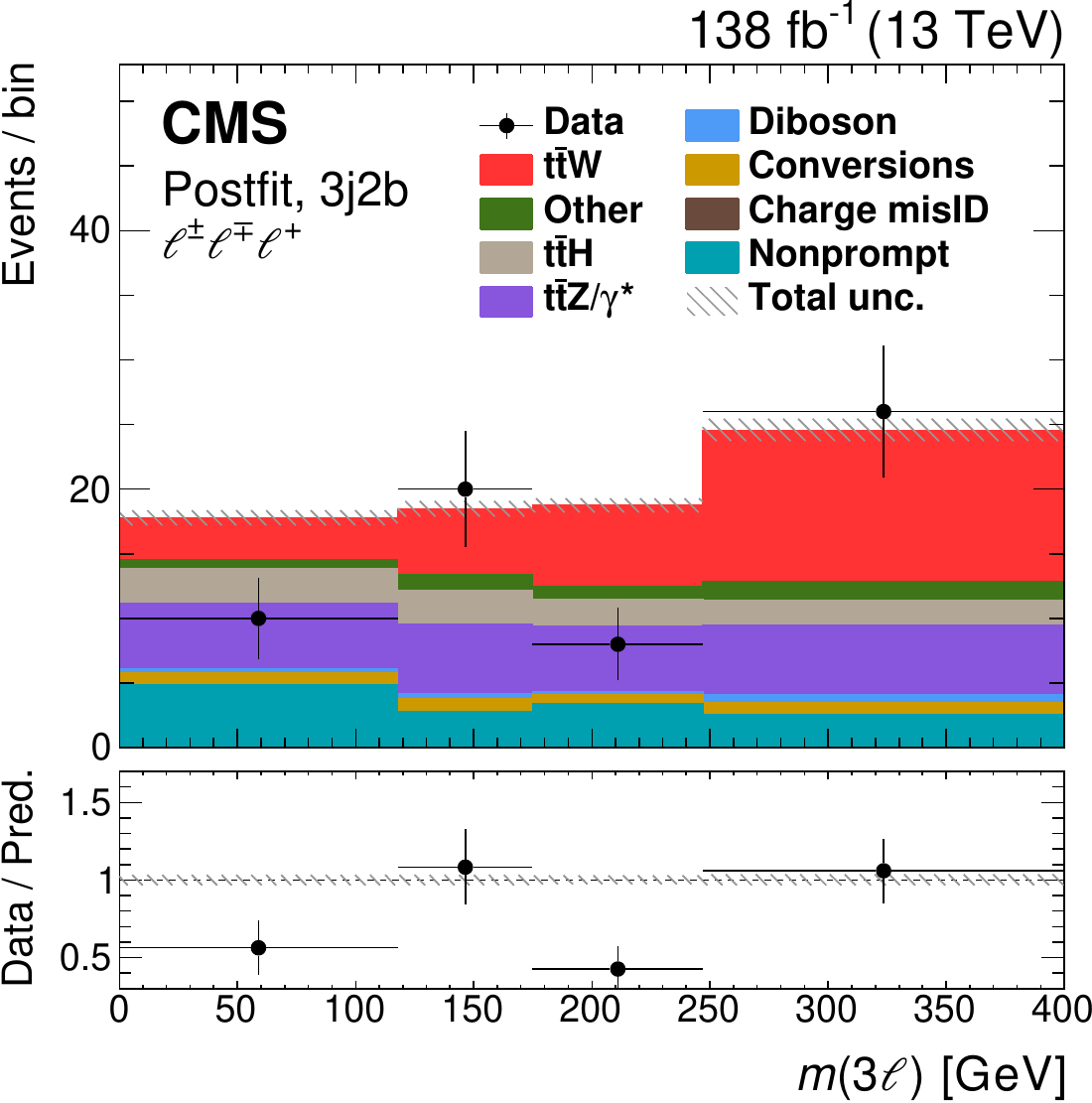}
\caption{%
    Comparison of the number of observed (points) and predicted (colored
    histograms) events in the trilepton signal region with positive sum of
    lepton charges.
    The distributions of the three-lepton invariant mass are displayed in final
    states with two (left) and three (right) jets, of which exactly two pass the
    medium \PQb tagging requirements.
    The predictions are shown ``prefit'' as in Fig.~\ref{fig:sel3l} (upper row),
    and with the values of the normalizations and nuisance parameters obtained
    in the fit to the data applied (``postfit'', lower row).
    The vertical bars on the points represent the statistical uncertainties in
    the data, the horizontal bars the bin widths, and the hatched bands the
    systematic uncertainty in the predictions.
    The last bins include the overflow contributions.
    In the lower panels, the ratio of the event yields in data to the overall
    sum of the predictions is presented.
}
\label{fig:fit3lpos}
\end{figure}

\begin{figure}[htp!]
\centering
\includegraphics[width=0.45\textwidth]{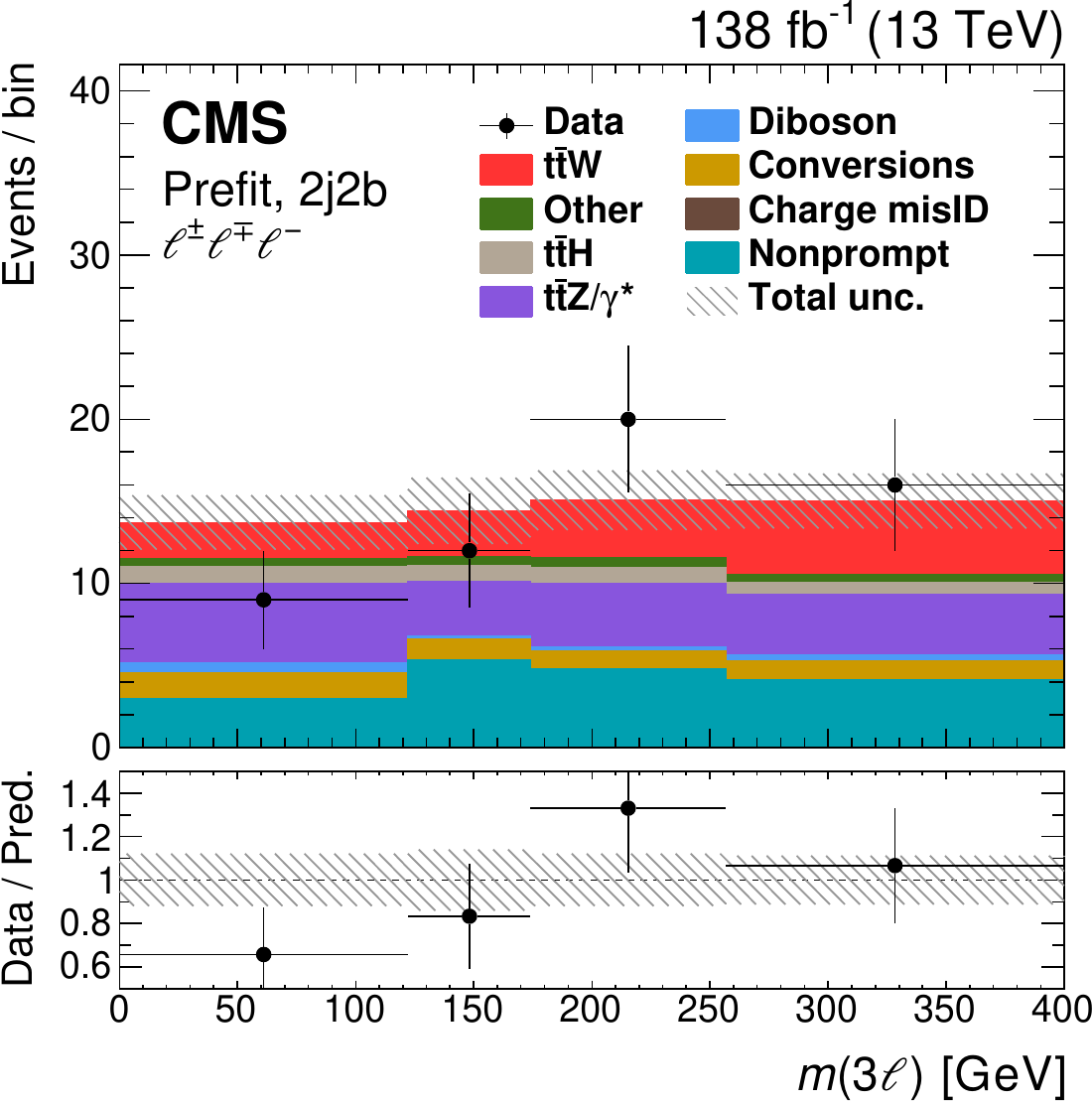}%
\hspace{0.05\textwidth}%
\includegraphics[width=0.45\textwidth]{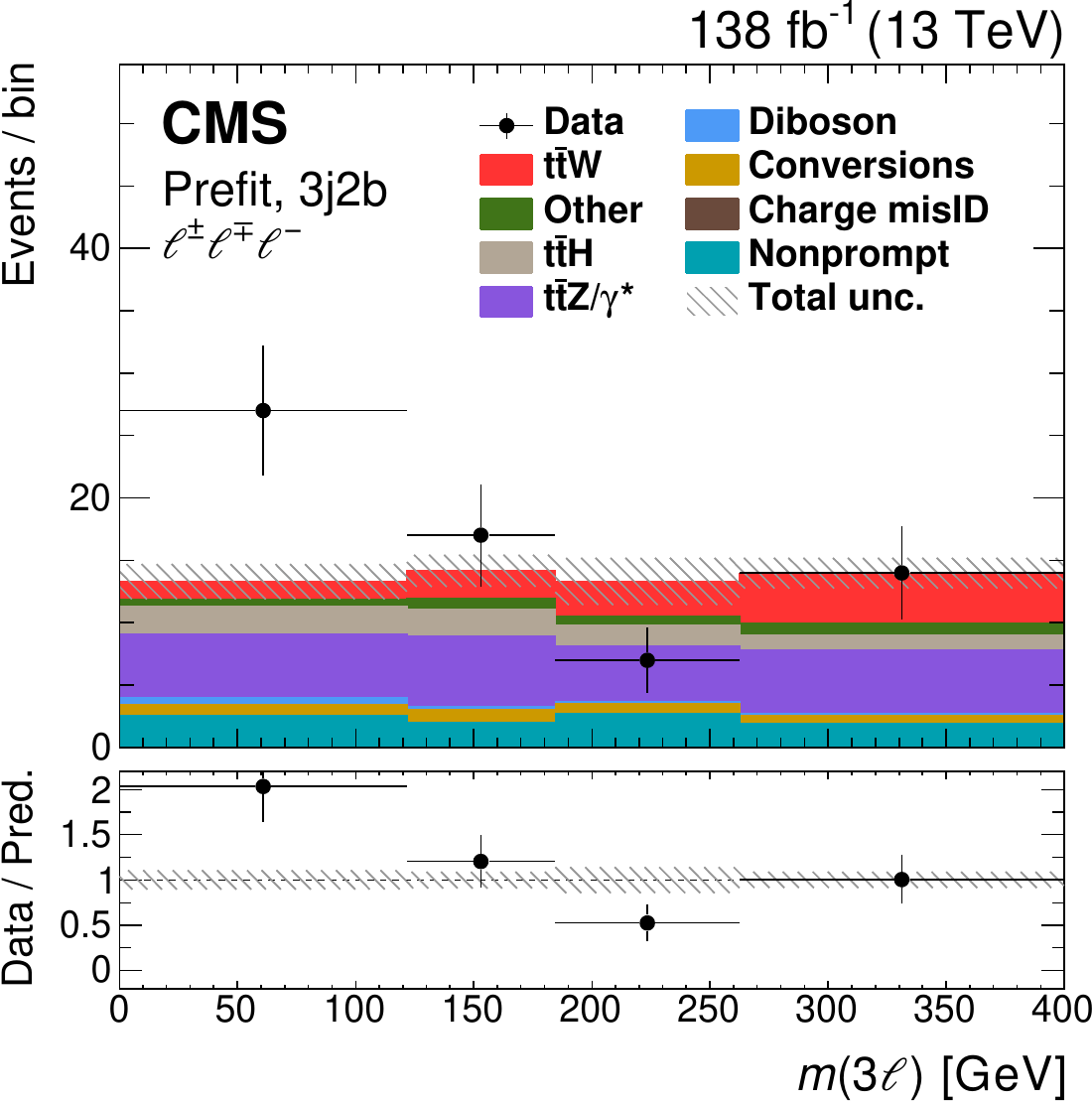} \\[1ex]
\includegraphics[width=0.45\textwidth]{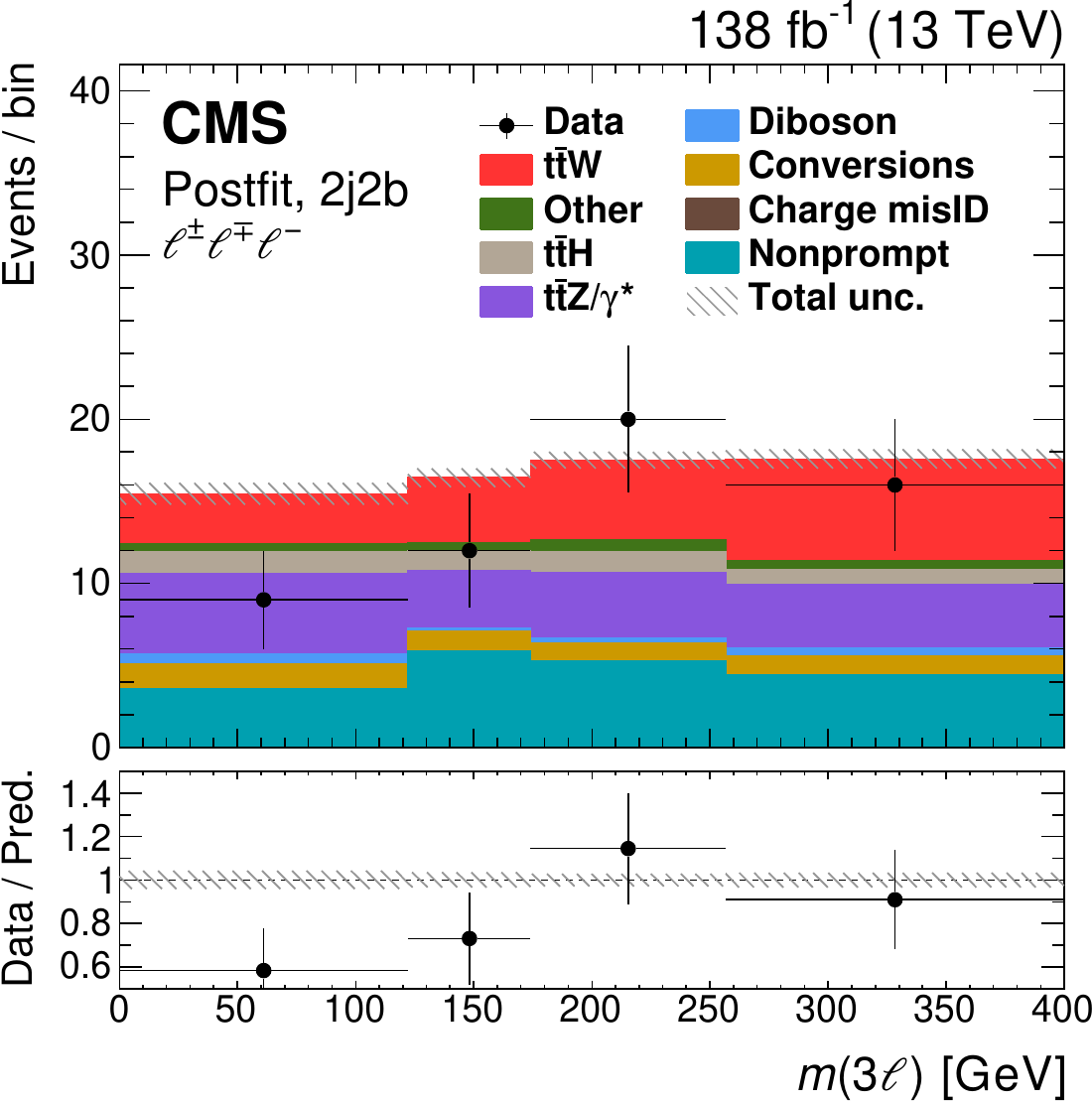}%
\hspace{0.05\textwidth}%
\includegraphics[width=0.45\textwidth]{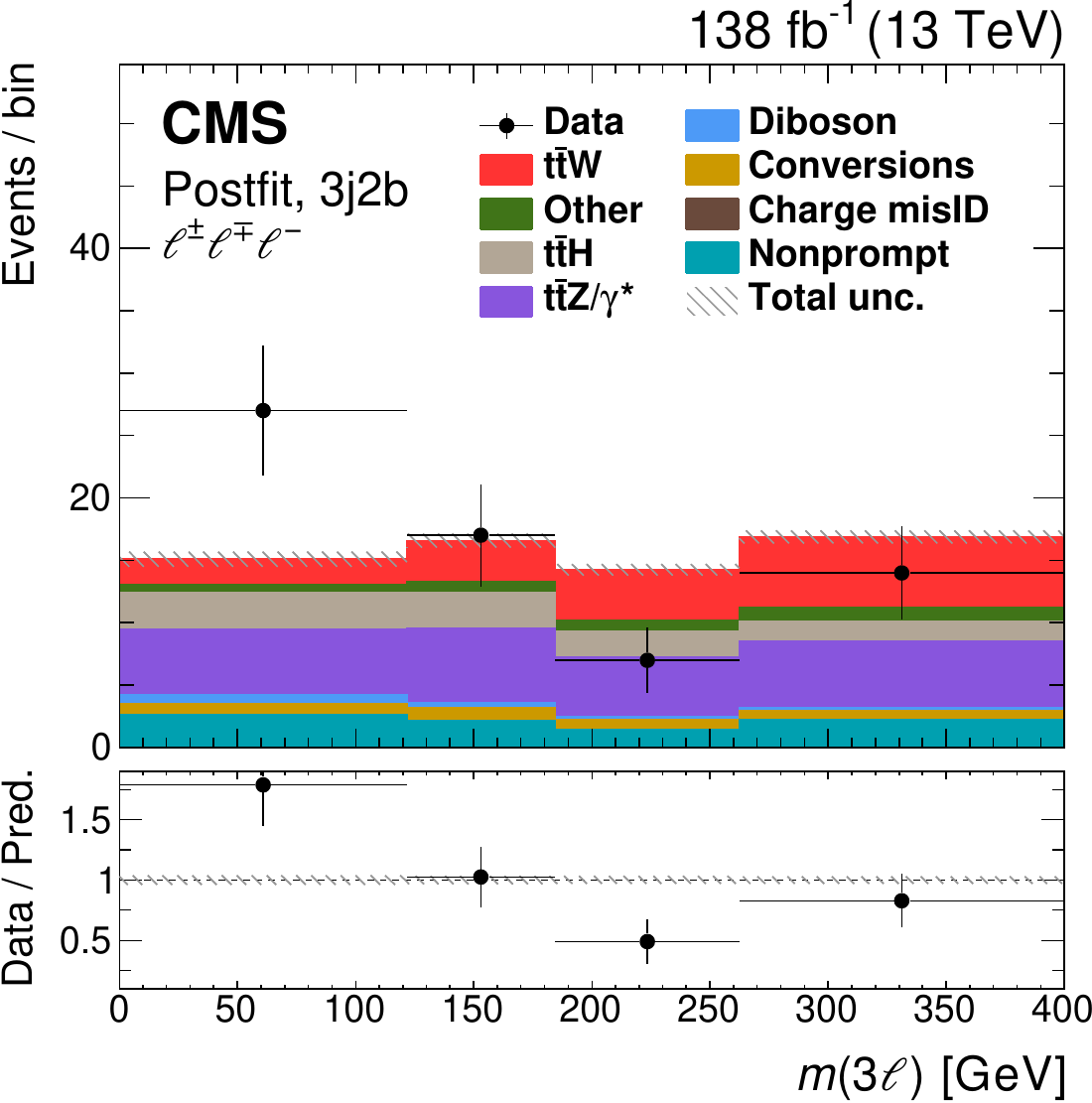}
\caption{%
    Comparison of the number of observed (points) and predicted (colored
    histograms) events in the trilepton signal region with negative sum of
    lepton charges.
    The distributions of the three-lepton invariant mass are displayed in final
    states with two (left) and three (right) jets, of which exactly two pass the
    medium \PQb tagging requirements.
    The predictions are shown ``prefit'' as in Fig.~\ref{fig:sel3l} (upper row),
    and with the values of the normalizations and nuisance parameters obtained
    in the fit to the data applied (``postfit'', lower row).
    The vertical bars on the points represent the statistical uncertainties in
    the data, the horizontal bars the bin widths, and the hatched bands the
    systematic uncertainty in the predictions.
    The last bins include the overflow contributions.
    In the lower panels, the ratio of the event yields in data to the overall
    sum of the predictions is presented.
}
\label{fig:fit3lneg}
\end{figure}

The observed and predicted NN discriminant distributions are shown in
Fig.~\ref{fig:fit2l} for events with positive (left) and negative (right)
charged leptons in the dilepton signal region, before (prefit, upper) and after
(postfit, lower) the fit to the data.
Since the lepton charges are input variables in the NN training, the NN is able
to separately optimize the separation of signal and background for events with
positive and negative lepton charges.
As a result, the shapes of the NN discriminant distributions are different for
the two classes of events.
Figures~\ref{fig:fit3lpos} and~\ref{fig:fit3lneg} show the \mlll distributions
for events in the trilepton signal region with total trilepton charge $+1$ and
$-1$, respectively.
The prefit (postfit) distributions are shown in the upper (lower) panels, with
events having 2 (left) and 3 (right) jets, of which exactly 2 are medium
\PQb-tagged jets, in both figures.
The $p$-value quantifying the agreement between the data and the prediction in
the postfit distributions of the trilepton signal region has a value of 0.027.

\begin{table}[!p]
\centering
\caption{%
    Sources of systematic uncertainty in the predicted signal and background
    event yields with their impact on the measured \ttW production cross
    section, estimated after the fit to the data.
    Only systematic uncertainty sources with values greater than 0.1\% are
    included in the table.
    The production cross sections of the \ttW, \WZ, \ZZ, and \ttZ processes are
    simultaneously constrained in the fit.
    The second-to-last row refers to the statistical uncertainty in the
    simulated event samples.
}
\renewcommand\arraystretch{1.2}
\begin{tabular}{p{1ex}l@{\hspace*{-1.5em}}c}
    \multicolumn{2}{l}{Source} & Uncertainty [\%] \\ \hline
    \multicolumn{2}{l}{Experimental uncertainties} & \\
    & Integrated luminosity & 1.9 \\
    & \PQb tagging efficiency & 1.6 \\
    & Trigger efficiency & 1.2 \\
    & Pileup reweighting & 1.0 \\
    & L1 inefficiency & 0.7 \\
    & Jet energy scale & 0.6 \\
    & Jet energy resolution & 0.4 \\
    & Lepton selection efficiency & 0.4 \\[\cmsTabSkip]
    \multicolumn{2}{l}{Background uncertainties} & \\
    & \ttH normalization & 2.6 \\
    & Charge misidentification & 1.6 \\
    & Nonprompt leptons & 1.3 \\
    & \VVV normalization & 1.2 \\
    & \ttVV normalization & 1.2 \\
    & Conversions normalization & 0.7 \\
    & \ttg normalization & 0.6 \\
    & \ZZ normalization  & 0.6 \\
    & Other normalizations & 0.5 \\
    & \ttZ normalization & 0.3 \\
    & \WZ normalization  & 0.2 \\
    & \tZq normalization & 0.2 \\
    & \tHq normalization & 0.2 \\[\cmsTabSkip]
    \multicolumn{2}{l}{Modeling uncertainties} & \\
    & \ttW scale & 1.8 \\
    & \ttW color reconnection & 1.0 \\
    & ISR \& FSR scale for \ttW & 0.8 \\
    & \ttg scale & 0.4 \\
    & \VVV scale & 0.3 \\
    & \ttH scale & 0.2 \\
    & Conversions & 0.2 \\[\cmsTabSkip]
    \multicolumn{2}{l}{Simulation statistical uncertainty} & 1.8 \\[\cmsTabSkip]
    \multicolumn{2}{l}{Total systematic uncertainty} & 5.8 \\
\end{tabular}
\label{tab:syst}
\end{table}

The sources and estimates of the various systematic uncertainties relevant to
this measurement are given in Table~\ref{tab:syst}.
The reduction in the uncertainties between the prefit and postfit values is
clearly evident in Figs.~\ref{fig:fit2l}--\ref{fig:fit3lneg}.
This is due to the categorization of the signal regions and the inclusion of the
various control regions in the fit, both of which help to constrain some of the
uncertainty sources.
The dominant experimental uncertainties include those in the integrated
luminosity measurement, the background estimation of electron charge
misidentification, and the \PQb jet identification efficiency.
The normalization uncertainties in the predictions of the \ttH, \VVV, and \ttVV
processes also represent substantial contributions to the total uncertainty in
the measurement.
Given the similarities of the decay products of these rare processes with those
of the \ttW signal process, they make a sizable contribution to the total
uncertainty despite their low number of excepted events.
Additional important sources arise from the modeling uncertainties associated
with the simulation of \ttW events and from the statistical uncertainty in the
predicted number of signal and background events.

Compared to the previous CMS measurement of \ttW
production~\cite{CMS:TOP-17-005}, which was performed with a subset of the data
analyzed here, we achieve a significant improvement in the precision of the
cross section measurement, with a reduction of the statistical (systematic)
uncertainty by a factor of three (more than two).
This is mainly attributed to the larger data sample, as well as to an optimized
analysis strategy and improved estimates of the dominant background
contributions using data control samples.
A higher selection efficiency of \ttW events and a stronger background
suppression of the nonprompt-lepton background contribution is achieved by the
lepton selection with the dedicated lepton MVA discriminant.
An additional source of improvement in the obtained signal-to-background ratio
is associated with the enhanced \PQb tagging performance following the
installation of the new pixel detector in 2017 and a more performant
\PQb tagging algorithm~\cite{CMS:BTV-16-002, EXT:DeepJet-2020, CMS:DP-2018-058}.

\begin{figure}[htb!]
\centering
\includegraphics[width=0.74\textwidth]{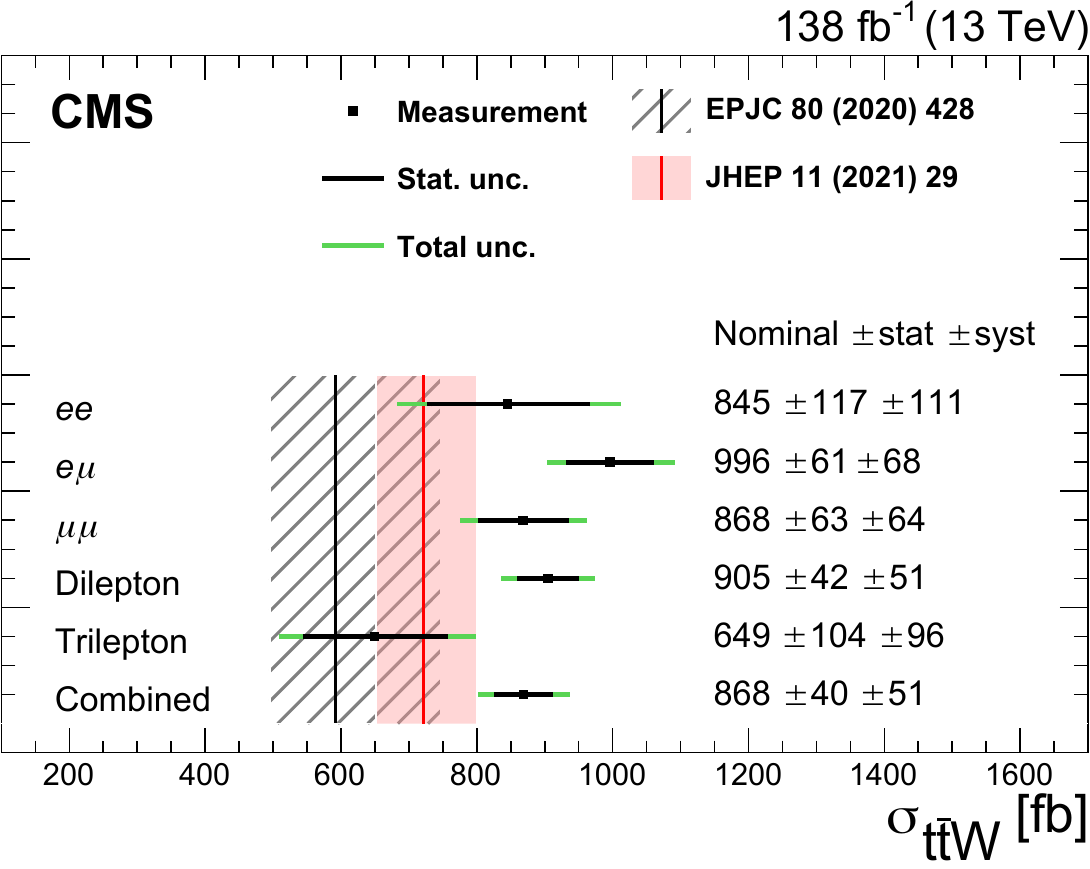}
\caption{%
    The measured \ttW production cross section for the individual dilepton
    ($\Pe\Pe$, $\Pe\PGm$, $\PGm\PGm$, and combined) channels and the trilepton
    channel, as well as their combination.
    The inner black bars show the statistical uncertainty, and the outer green
    bars give the total uncertainty.
    The predictions from two SM calculations from Refs.~\cite{Kulesza:2020nfh,
    Frederix:2021agh} are shown by the black and red vertical lines, with the
    associated bands corresponding to the total uncertainty.
}
\label{fig:xsecincl}
\end{figure}

The measured \ttW inclusive cross section is shown in Table~\ref{tab:results}
with its complete statistical and systematic uncertainties.
A comparison of the measured \sigTTW values obtained from this analysis using
different final states separately is shown in Fig.~\ref{fig:xsecincl}.
In addition, two SM predictions are shown both in Table~\ref{tab:results} and
Fig.~\ref{fig:xsecincl}.
In comparison both with the SM \NLONNLL calculation~\cite{Kulesza:2020nfh} and
the SM prediction using NLO plus an improved FxFx merging
procedure~\cite{Frederix:2021agh}, the measured cross section is larger but
consistent within two standard deviations.
The constraints on \ttW production obtained in the \ttH and \tttt studies in
Refs.~\cite{CMS:HIG-19-008, CMS:TOP-18-003} are consistent with our measured
cross section within one standard deviation, and confirm the underprediction of
the \ttW cross section by the SM predictions.

\begin{table}[!htp]
\centering\renewcommand\arraystretch{1.4}
\topcaption{%
    Summary of measured and predicted production cross sections of \ttW, \ttWp,
    and \ttWm production, as well as of the \RTTWpm ratio.
    The SM predictions quoted at \NLONNLL accuracy are taken from
    Refs.~\cite{Kulesza:2020nfh, Broggio:2019ewu}.
    The SM predictions quoted at NLO accuracy and including corrections from an
    improved FxFx merging procedure (\NLOFXFX) have been provided by the authors
    of Ref.~\cite{Frederix:2021agh}.
    The theoretical uncertainties include scale variations and PDF
    uncertainties.
}
\begin{tabular}{ccccc}
    \multirow{2}{*}{Observable} & \multirow{2}{*}{Measurement} & \multicolumn{2}{c}{SM prediction} \\[-4pt]
    & & \NLONNLL & \NLOFXFX \\ \hline
    \sigTTW & \ExpXsecTTW & \PredXsecTTWnnlo & $722\,^{+71}_{-78}\thy\fb$ \\
    \sigTTWp & \ExpXsecTTWp & $384\,^{+53}_{-33}\thy\fb$ & $475\,^{+46}_{-52}\thy\fb$ \\
    \sigTTWm & \ExpXsecTTWm & $198\,^{+26}_{-17}\thy\fb$ & $247\,^{+24}_{-27}\thy\fb$ \\
    \RTTWpm & \ExpXsecTTWr & $1.94\,^{+0.37}_{-0.24}\thy$ & $1.92\,^{+0.27}_{-0.29}\thy$ \\
\end{tabular}
\label{tab:results}
\end{table}

\begin{figure}[htbp!]
\centering
\includegraphics[width=0.55\textwidth]{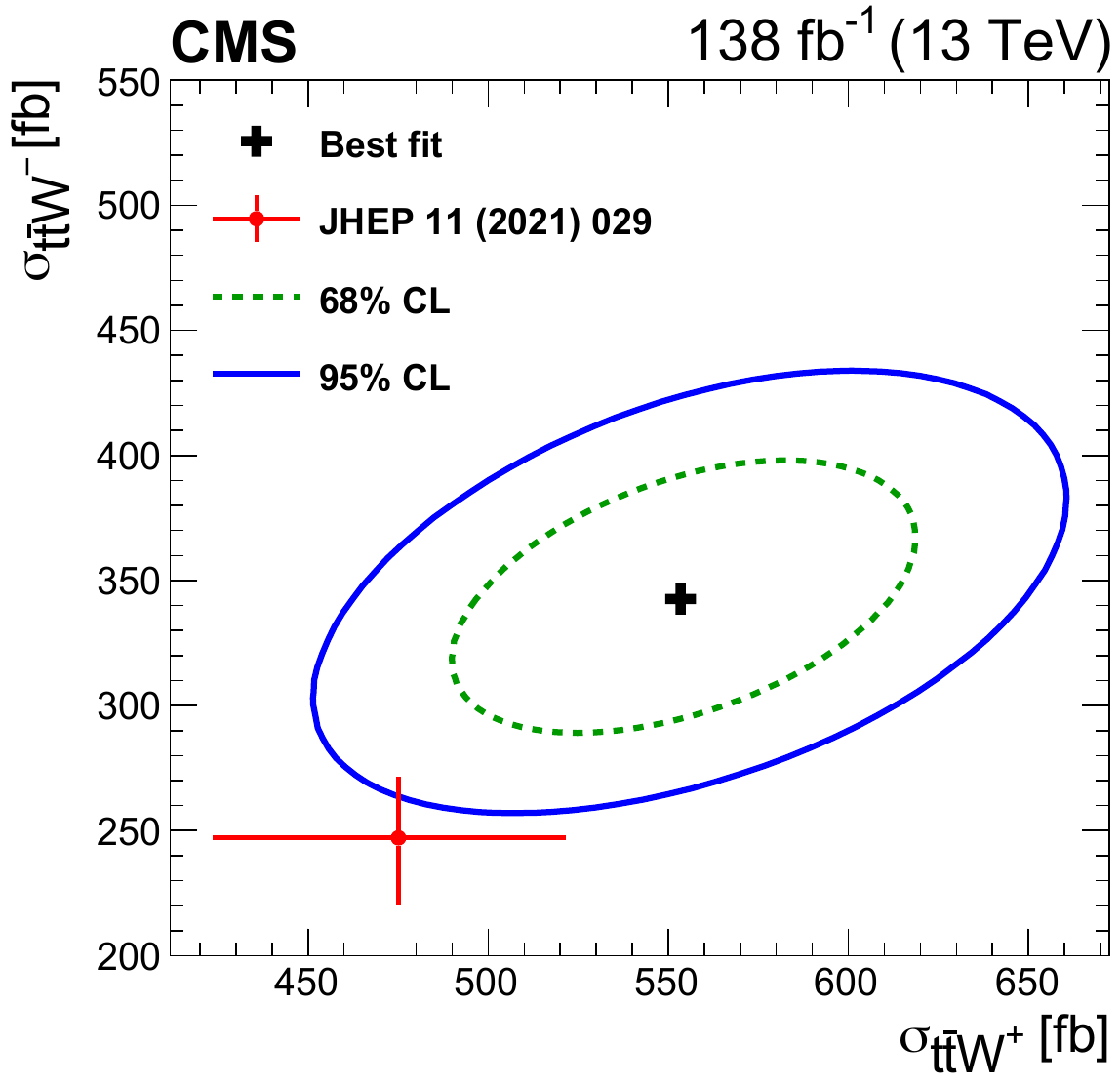}
\caption{%
    The measurement of the \ttWp vs. \ttWm cross sections (black cross), along
    with the 68 (green dashed) and 95\% (blue solid) \CL intervals.
    The SM prediction provided by the authors of Ref.~\cite{Frederix:2021agh} is
    shown by the red dot, with the horizontal and vertical bars corresponding to
    the total uncertainties.
}
\label{fig:xsecpm}
\end{figure}

The separate \ttWp and \ttWm cross section measurements are of special interest
because of the enhancement in \pp collisions of the former with respect to the
latter due to valence quark effects, and therefore provide important information
on the PDF of the proton.
The cross sections are measured in a simultaneous fit to data, where the events
in categories with positive (negative) sum of lepton charges are used to
constrain the \ttWp (\ttWm) cross section.
The resulting cross sections and corresponding uncertainties are given in
Table~\ref{tab:results} and shown in Fig.~\ref{fig:xsecpm}.
The correlation between the two measured cross sections originates from the use
of common nuisance parameters and the same control regions in the fit.
The sum of the two measured cross sections is consistent with the measured
inclusive \ttW cross section within the statistical uncertainties.
Both cross sections are larger than but consistent with the SM \NLONNLL
calculations from Ref.~\cite{Broggio:2019ewu} and the SM NLO predictions using
an improved FxFx merging procedure provided by the authors of
Ref.~\cite{Frederix:2021agh}.
These predictions and their uncertainties are shown in Table~\ref{tab:results}.

\begin{figure}[htbp!]
\centering
\includegraphics[width=0.6\textwidth]{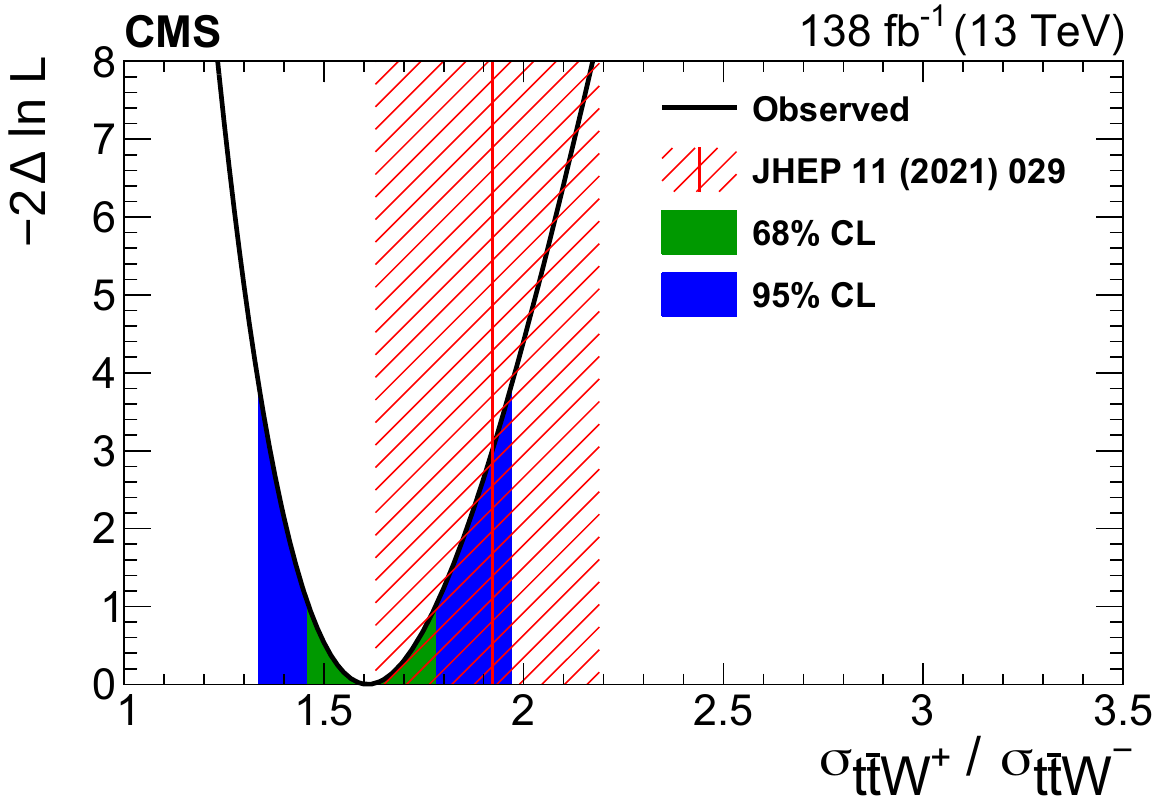}
\caption{%
    The scan of the negative log-likelihood used in the measurement of the cross
    section ratio \RTTWpm.
    The best fit value is indicated, with the inner green and outer blue bands
    displaying the 68 and 95\% \CL intervals, respectively.
    An SM prediction and its uncertainty, found from the cross section values
    provided by the authors of Ref.~\cite{Frederix:2021agh}, are shown by the
    vertical red line and hatched red band.
}
\label{fig:ratio}
\end{figure}

The measurement of the ratio between the cross sections of the two processes,
\RTTWpm, is performed by modifying the fit procedure to directly measure the
ratio.
Figure~\ref{fig:ratio} shows the scan of the negative
log-likelihood~\cite{EXT:Asymptotic-2011}, with the accompanying best fit value
and 68 and 95\% confidence level (\CL) intervals for the ratio.
The ratio measurement has a reduced relative systematic uncertainty compared to
the individual cross sections because of partial correlations between some of
the uncertainty sources that cancel in the ratio.
The measured ratio is smaller than the ratios calculated from the theory
predictions~\cite{Broggio:2019ewu, Frederix:2021agh}, all of which are shown in
Table~\ref{tab:results}.
In this study, we have propagated the uncertainties in the SM prediction of the
\ttWp and \ttWm cross sections to the uncertainty in the ratio assuming no
correlation, which results in an overestimation of the latter uncertainty.

\section{Summary}
\label{sec:summary}

The cross section of the associated production of a \PW boson with a top
quark-antiquark pair (\ttW) in proton-proton collisions at \sqrts is measured
using data collected with the CMS detector in 2016--2018, corresponding to an
integrated luminosity of \il.
The measurement is performed in events with two same-sign or three charged
leptons (electrons and muons) and additional jets.
Events are categorized according to the charges and flavors of the leptons, and
the number of jets and jets from the hadronization of \PQb quarks.
In the dilepton channel, a multiclass neural network is used to improve the
separation of signal from background processes.
In the trilepton channel, the invariant mass of the three leptons is used as the
discriminating variable.
Background contributions with leptons originating from hadron decays, jets or
hadrons misidentified as leptons, and with misreconstructed charge are estimated
from control samples in data.
All other background processes are estimated from simulated event samples, and
control samples in data are used to constrain the dominant contributions.

The inclusive \ttW production cross section in the full phase space is measured
to be \ExpXsecTTW.
The \ttWp and \ttWm cross sections are also determined as \ExpXsecTTWpNU and
\ExpXsecTTWm, respectively.
The measured ratio between these production cross sections is \ExpXsecTTWr.
The measured cross sections are larger than but consistent with standard model
predictions within two standard deviations, while the \RTTWpm ratio is smaller
than but consistent with the predicted value.
These results are all consistent with earlier measurements and are the most
precise determinations published to date.

\begin{acknowledgments}
We congratulate our colleagues in the CERN accelerator departments for the excellent performance of the LHC and thank the technical and administrative staffs at CERN and at other CMS institutes for their contributions to the success of the CMS effort. In addition, we gratefully acknowledge the computing centers and personnel of the Worldwide LHC Computing Grid and other centers for delivering so effectively the computing infrastructure essential to our analyses. Finally, we acknowledge the enduring support for the construction and operation of the LHC, the CMS detector, and the supporting computing infrastructure provided by the following funding agencies: BMBWF and FWF (Austria); FNRS and FWO (Belgium); CNPq, CAPES, FAPERJ, FAPERGS, and FAPESP (Brazil); MES and BNSF (Bulgaria); CERN; CAS, MoST, and NSFC (China); MINCIENCIAS (Colombia); MSES and CSF (Croatia); RIF (Cyprus); SENESCYT (Ecuador); MoER, ERC PUT and ERDF (Estonia); Academy of Finland, MEC, and HIP (Finland); CEA and CNRS/IN2P3 (France); BMBF, DFG, and HGF (Germany); GSRI (Greece); NKFIH (Hungary); DAE and DST (India); IPM (Iran); SFI (Ireland); INFN (Italy); MSIP and NRF (Republic of Korea); MES (Latvia); LAS (Lithuania); MOE and UM (Malaysia); BUAP, CINVESTAV, CONACYT, LNS, SEP, and UASLP-FAI (Mexico); MOS (Montenegro); MBIE (New Zealand); PAEC (Pakistan); MES and NSC (Poland); FCT (Portugal); MESTD (Serbia); MCIN/AEI and PCTI (Spain); MOSTR (Sri Lanka); Swiss Funding Agencies (Switzerland); MST (Taipei); MHESI and NSTDA (Thailand); TUBITAK and TENMAK (Turkey); NASU (Ukraine); STFC (United Kingdom); DOE and NSF (USA).
    
\hyphenation{Rachada-pisek} Individuals have received support from the Marie-Curie program and the European Research Council and Horizon 2020 Grant, contract Nos.\ 675440, 724704, 752730, 758316, 765710, 824093, 884104, and COST Action CA16108 (European Union); the Leventis Foundation; the Alfred P.\ Sloan Foundation; the Alexander von Humboldt Foundation; the Belgian Federal Science Policy Office; the Fonds pour la Formation \`a la Recherche dans l'Industrie et dans l'Agriculture (FRIA-Belgium); the Agentschap voor Innovatie door Wetenschap en Technologie (IWT-Belgium); the F.R.S.-FNRS and FWO (Belgium) under the ``Excellence of Science -- EOS" -- be.h project n.\ 30820817; the Beijing Municipal Science \& Technology Commission, No. Z191100007219010; the Ministry of Education, Youth and Sports (MEYS) of the Czech Republic; the Hellenic Foundation for Research and Innovation (HFRI), Project Number 2288 (Greece); the Deutsche Forschungsgemeinschaft (DFG), under Germany's Excellence Strategy -- EXC 2121 ``Quantum Universe" -- 390833306, and under project number 400140256 - GRK2497; the Hungarian Academy of Sciences, the New National Excellence Program - \'UNKP, the NKFIH research grants K 124845, K 124850, K 128713, K 128786, K 129058, K 131991, K 133046, K 138136, K 143460, K 143477, 2020-2.2.1-ED-2021-00181, and TKP2021-NKTA-64 (Hungary); the Council of Science and Industrial Research, India; the Latvian Council of Science; the Ministry of Education and Science, project no. 2022/WK/14, and the National Science Center, contracts Opus 2021/41/B/ST2/01369 and 2021/43/B/ST2/01552 (Poland); the Funda\c{c}\~ao para a Ci\^encia e a Tecnologia, grant CEECIND/01334/2018 (Portugal); the National Priorities Research Program by Qatar National Research Fund; MCIN/AEI/10.13039/501100011033, ERDF ``a way of making Europe", and the Programa Estatal de Fomento de la Investigaci{\'o}n Cient{\'i}fica y T{\'e}cnica de Excelencia Mar\'{\i}a de Maeztu, grant MDM-2017-0765 and Programa Severo Ochoa del Principado de Asturias (Spain); the Chulalongkorn Academic into Its 2nd Century Project Advancement Project, and the National Science, Research and Innovation Fund via the Program Management Unit for Human Resources \& Institutional Development, Research and Innovation, grant B05F650021 (Thailand); the Kavli Foundation; the Nvidia Corporation; the SuperMicro Corporation; the Welch Foundation, contract C-1845; and the Weston Havens Foundation (USA).
\end{acknowledgments}

\bibliography{auto_generated}
\cleardoublepage \appendix\section{The CMS Collaboration \label{app:collab}}\begin{sloppypar}\hyphenpenalty=5000\widowpenalty=500\clubpenalty=5000
\cmsinstitute{Yerevan Physics Institute, Yerevan, Armenia}
{\tolerance=6000
A.~Tumasyan\cmsAuthorMark{1}\cmsorcid{0009-0000-0684-6742}
\par}
\cmsinstitute{Institut f\"{u}r Hochenergiephysik, Vienna, Austria}
{\tolerance=6000
W.~Adam\cmsorcid{0000-0001-9099-4341}, J.W.~Andrejkovic, T.~Bergauer\cmsorcid{0000-0002-5786-0293}, S.~Chatterjee\cmsorcid{0000-0003-2660-0349}, K.~Damanakis\cmsorcid{0000-0001-5389-2872}, M.~Dragicevic\cmsorcid{0000-0003-1967-6783}, A.~Escalante~Del~Valle\cmsorcid{0000-0002-9702-6359}, P.S.~Hussain\cmsorcid{0000-0002-4825-5278}, M.~Jeitler\cmsAuthorMark{2}\cmsorcid{0000-0002-5141-9560}, N.~Krammer\cmsorcid{0000-0002-0548-0985}, L.~Lechner\cmsorcid{0000-0002-3065-1141}, D.~Liko\cmsorcid{0000-0002-3380-473X}, I.~Mikulec\cmsorcid{0000-0003-0385-2746}, P.~Paulitsch, F.M.~Pitters, J.~Schieck\cmsAuthorMark{2}\cmsorcid{0000-0002-1058-8093}, R.~Sch\"{o}fbeck\cmsorcid{0000-0002-2332-8784}, D.~Schwarz\cmsorcid{0000-0002-3821-7331}, S.~Templ\cmsorcid{0000-0003-3137-5692}, W.~Waltenberger\cmsorcid{0000-0002-6215-7228}, C.-E.~Wulz\cmsAuthorMark{2}\cmsorcid{0000-0001-9226-5812}
% ...
\par}
\vskip\cmsinstskip
\dag:~Deceased\\
$^{1}$Also at Yerevan State University, Yerevan, Armenia\\
$^{2}$Also at TU Wien, Vienna, Austria\\
% ...
\end{sloppypar}
\end{document}